\documentstyle[preprint,aps]{revtex}
\begin{document}
\draft
\preprint{
\vbox{
\halign{&##\hfil\cr
	& AS-ITP-98-09 \cr
        & hep-ph/9808379 \cr
	& Revised in Mar. 1999 \cr}}
}
\title{Rare B Decays in Supersymmetry without R-parity} 
\author{Tai-fu Feng \footnote{email: fengtf@itp.ac.cn}}
\address{Institute of Theoretical Physics, Academia Sinica, P. O. Box 2735,
Beijing 100080, P.R.China}
\maketitle
\begin{center}
\begin{abstract}
We perform the complete computation of the contributions to $b\rightarrow 
 s+\gamma$, $b\rightarrow s e^{+}e^{-}$, $b\rightarrow s\sum \nu_{i} \bar{\nu}
_{i}$ in supersymmetric model with bilinear R-parity violation. We compare our calculations with
the evaluations in SM and the experimental results. We find that the supersymmetric contributions
can be quite large in those processes. From the analysis and experimental results, we can get some
constraints on the mass spectrum in the model.
\end{abstract}
\end{center}
\section{Introduction}

It is being increasingly realized by those engaged in the search for the
supersymmetry (SUSY)\cite{s1} that the principle of R-parity conservation,
 assumed to be sacrosanct in the prevalent search strategies, is not inviolable
in practice. The R-parity of a particle is defined as $ R=(-1)^{L+3B+2S}$, and 
can be violated if either baryon (B) or lepton (L) number is 
not conserved in nature, a fact perfectly compatible with the non observation
of proton decay. Under R-parity violation the phenomenology changes considerably\cite{s2,s3}, 
the most important consequence is that the lightest 
supersymmetric particle (LSP) can decay now. However, the way in which R-parity 
can be violated is not unique. Different types of R-parity violating
interaction terms can be written down, leading to different observable
predictions. In addition, R-parity can be violated spontaneously, instead
of explicitly, whence another class of interesting effects are expected
\cite{s4}. If the phenomenology of R-parity breaking has to be 
understood, and the consequent modifications in the current search strategies
have to be effectively implemented, then it is quite important to explore
the full implication of each possible R-breaking scheme.

The R-conserving part of the minimal supersymmetric standard model (MSSM) is
specified by the superpotential
\begin{eqnarray}
{\cal W}_{MSSM} & = & \mu \varepsilon_{ij} \hat{H}_{i}^{1} \hat{H}_{j}^{2} +
\varepsilon_{ij} l_{IJ} \hat{H}_{i}^{1} \hat{L}_{j}^{I} \hat{R}^{J} +
\varepsilon_{ij} d_{IJ} \hat{H}_{i}^{1} \hat{Q}_{j}^{I} \hat{D}^{J}   \nonumber 
\\
  &  & +\varepsilon_{ij} u_{IJ} \hat{H}_{i}^{2} \hat{Q}_{j}^{I} \hat{U}^{J} , 
\end{eqnarray}
where $I$, $J = 1$, $2$, $3$ are generation indices; $i$, $j = 1$, $2$ are SU(2)
indices; and $ \varepsilon $ is a completely antisymmetric $ 2\times 2$
matrix, with $ \varepsilon_{12} = 1$. The $"$hat$"$ symbol over each letter
indicates a superfield, $\hat{Q}^{I}$, $\hat{L}^{I}$, $\hat{H}^{1}$, and $
\hat{H}^{2} $ being SU(2) doublets with hyper-charges $ \frac{1}{3}$, $-1$,
$-1$, and $1$ respectively; $\hat{U}$, $\hat{D}$ and $\hat{R} $ being SU(2)
singlets with hyper-charges $-\frac{4}{3}$, $\frac{2}{3} $ and 2 respectively.
The couplings $u_{IJ}$, $d_{IJ}$ and $l_{IJ}$ are $3 \times 3$ Yukawa
matrices, and $\mu$ is parameter with unit of mass.
If now the bilinear R-parity violating interactions are incorporated, the superpotential
takes the form\cite{s5} 
\begin{equation}
{\cal W}={\cal W}_{MSSM} + {\cal W}_{L}
\label{rpvsuperpoten}
\end{equation}
with $ {\cal W}_{L} = \varepsilon_{ij} \epsilon^{\prime}_{I}\hat{H}_{i}^{2}\hat{L}_{j}
^{I} $ and $\epsilon^{\prime}_{I}$ is the parameter with unit of mass.
  
The soft SUSY-breaking terms:
\begin{eqnarray}
{\cal L}_{soft} &=& -m_{H^{1}}^{2} H_{i}^{1*} H_{i}^{1}-m_{H^{2}}^{2} H_{i}^{2*}
H_{i}^{2}-m_{L^{I}}^{2} \tilde{L}_{i}^{I*} \tilde{L}_{i}^{I}  \nonumber  \\
  & & -m_{R^{I}}^{2} \tilde{R}^{I*} \tilde{R}^{I}-m_{Q^{I}}^{2} \tilde{Q}_{i}^{I*}
\tilde{Q}_{i}^{I}-m_{D^{I}}^{2} \tilde{D}^{I*} \tilde{D}^{I}  \nonumber  \\
 & &-m_{U^{I}}^{2} \tilde{U}^{I*} \tilde{U}^{I} + (m_{1} \lambda_{B} \lambda_{B}
 + m_{2} \lambda_{A}^{i} \lambda_{A}^{i}  \nonumber  \\
 & & + m_{3} \lambda_{G}^{a} \lambda_{G}^{a} + h.c.) + (B \mu \varepsilon_
{ij} H_{i}^{1} H_{j}^{2} + B_{I} \epsilon^{\prime}_{I} \varepsilon_{ij} H_{i}^{2}
\tilde{L}_{j}^{I}  \nonumber  \\
 & & + \varepsilon_{ij} l_{sI} \mu H_{i}^{1} \tilde{L}_{j}^{I} \tilde{R}^{I}
+\varepsilon_{ij} d_{sI} \mu H_{i}^{1} \tilde{Q}_{j}^{I} \tilde{D}^{I}  
 \nonumber  \\
 & & + \varepsilon_{ij} u_{sI} \mu H_{i}^{2} \tilde{Q}_{j}^{I} \tilde{U}^{I}
 + h.c.)
\end{eqnarray}
where $m_{H^{1}}^{2}$, $m_{H^{2}}^{2}$, $m_{L^{I}}^{2}$, $m_{R^{I}}^{2}$, $m_{Q^{I}}^{2}$, 
$m_{D^{I}}^{2}$
and $m_{U^{I}}^{2}$ stand for the mass squared of the scalar fields while $m_{1}$, $m_{2}$, 
$m_{3}$ denote the masses of the ${\rm SU}(3)\times {\rm SU}(2) \times {\rm U}(1)$ gauginos
$\lambda_{G}^{a}$, $\lambda_{A}^{i}$ and $\lambda_{B}$, $B$ and $B_{I}$ ($I = 1$, $2$, $3$)
are free parameters with unit of mass.

${\cal W}_{L}$ and lepton-number breaking terms in ${\cal L}_{soft}$ give a
viable agent for R-parity breaking. It is particularly interesting for the fact
that it can trigger a mixing between neutralinos and neutrinos as well as between
charginos and charged leptons, resulting in observable effects. 
For $\epsilon^{\prime}_{3}$ can reach a large value (such as $|\epsilon^{\prime}_{3}| \sim
500 {\rm GeV}$) even though we have considered the experimental fact that $m_{\tau} = 1.77$GeV 
and $m_{\nu_{\tau}} \leq 24$MeV\cite{s5}, the mixing between charginos and charged leptons may give
an important contribution to the rare B decays such as $b\rightarrow s+\gamma$,
$b\rightarrow se^{+}e^{-}$, $b\rightarrow s\sum \nu_{i}\bar{\nu}_{i}$.

On the other hand, flavor-changing neutral current (FCNC) processes have constantly
played a major role in the development of electroweak theories. FCNC studies have 
found kaon physics as their best research ground so far. However, the achievements
in B-physics make B-mesons the new challenging frontier in the study of FCNC
phenomena. At least some of the rare B-processes which are induced by the FCNC
transitions $b\rightarrow s$ and $b\rightarrow d$ are within the reach of present
machines. Needless to say, the test of such rare B-processes presents a probe of
the validity of crucial ingredients of the Standard Model (SM) and, possibly of
the existence of new physics beyond the SM at the low energy scale.

$b\rightarrow s+\gamma $ and $b\rightarrow se^{+}e^{-}$ in the SM are analyzed
in  Ref\cite{s6,s7}. The SM prediction for the branching ratio (Br) of the 
inclusive decay $b\rightarrow s+\gamma $, once the large QCD corrections are included
\cite{s6}, is a few times $10^{-4}$. As for the semileptonic $b\rightarrow s
e^{+}e^{-}$ decay, the QCD- corrected SM prediction for its Br is about $10^{-6}$
\cite{s7}.

FCNC rare B-processes have been widely analyzed as a potential probe for extensions
of the SM implying new physics at the TeV scale. In the model with Two Higgs Doublets 
(THDM) and no tree level FCNC, the rare B decays and $B_{d} - \bar{B}_{d} $ mixing are
computed in Ref$\cite{s8,s9}$. As for left-right symmetric models, the simplest 
choice of taking the CKM mixing angles in the right-handed sector equal to the corresponding
left-handed ones (manifest or pseudo-manifest left-right symmetry) does not give any
appreciable effects with respect to the SM estimates of rare B-processes\cite{s10}.
Analogous results are obtained in the general case if no fine tunings of the parameters
are allowed\cite{s11}.

FCNC in the supersymmetry with R-parity has been discussed widely in Ref\cite{s12},
Ref\cite{s13,s14,s15}, FCNC in supersymmetry without R-parity has been discussed in Ref\cite{s16},
the review of FCNC in the supersymmetric models can be found in Ref\cite{s17}.

This paper is organized as follows.  In sect. II, we give a description of the general structure
of the supersymmetry with bilinear R-parity violation (BRPV). In sect. III, we provide the complete
analyses to  $b \rightarrow s+\gamma $, $b\rightarrow se^{+}e^{-}$, $b \rightarrow s\sum\nu_{i}
\bar{\nu}_{i}$ in supersymmetric model with bilinear R-parity violation. In sect. IV, we perform a numerical
analysis of those processes and compare them with the SM predictions and the experimental 
results. For completeness,
we give the analytic expression for Feynman integrals which one encounters in the 
evaluation of the amplitudes listed in sect. III in appendix. A. A systematic notation for the 
relevant Feynman rules involving R-parity breaking terms are introduced in appendix. B. 
In appendix. C, we give the mass matrix of $\tau$ neutrino-neutralinos. For
simplicity, we neglect the generation mixing of slepton and squark in the following 
analyses.

\section{Minimal SUSY Model with Bilinear R-parity violation}

From the superpotential Eq.\ (\ref{rpvsuperpoten}), we can perform an operation
that is the same as in the standard model by redefinition of the fields\cite{s24}
\begin{eqnarray}
 \hat{Q}_{i}^{I} & \rightarrow & V_{Q_{i}}^{IJ} \hat{Q}_{i}^{J},  \nonumber \\
 \hat{U}^{I}  &  \rightarrow & V_{U}^{IJ} \hat{U}^{J} , \nonumber \\
 \hat{D}^{I}  &  \rightarrow & V_{D}^{IJ} \hat{D}^{J} , \nonumber \\
 \hat{L}_{i}^{I} & \rightarrow & V_{L_{i}}^{IJ} \hat{L}_{i}^{J},  \nonumber \\
 \hat{R}^{I}  &  \rightarrow & V_{R}^{IJ} \hat{R}^{J} .
 \label{redefin}
\end{eqnarray}
One can diagonalize the matrices $l_{IJ}$, $u_{IJ}$ and $ d_{IJ} $, the
superpotential has the form
\begin{eqnarray}
{\cal W} & = & \mu \varepsilon_{ij} \hat{H}_{i}^{1} \hat{H}_{j}^{2} + l_{I}
\varepsilon_{ij} \hat{H}_{i}^{1} \hat{L}_{j}^{I} \hat{R}^{I} - 
u_{I}(\hat{H}^{2}_{1}
C^{JI*} \hat{Q}^{J}_{2}  \nonumber  \\
 & & - \hat{H}_{2}^{2} \hat{Q}^{I}_{1})\hat{U}^{I} - d_{I}(\hat{H}_{1}^{1} 
\hat{Q}
_{2}^{I} - \hat{H}_{2}^{1} C^{IJ}\hat{Q}_{1}^{J})\hat{D}^{I}  \nonumber  \\
 & & + \epsilon_{I} \varepsilon_{ij} \hat{H}_{i}^{2} \hat{L}_{j}^{I}  ,\label{supernew}
\end{eqnarray}
where C is the Kobayashi-Maskawa matrix and $\epsilon_{I}$ ($I = 1$, $2$, $3$)
have the definition as:
\begin{eqnarray}
  C & = & V_{Q_{2}}^{\dag} V_{Q_{1}}  ,\nonumber \\
  \epsilon_{I} & = & \epsilon^{\prime}_{J} V_{L}^{JI} .
  \label{eq-5}
\end{eqnarray}
The soft breaking sector has the form
\begin{eqnarray}
{\cal L}_{soft} & = & -m_{H^{1}}^{2}H_{i}^{1*}H_{i}^{1} - m_{H^{2}}^{2} H_{i}^{2*}
H_{i}^{2}-m_{L^{I}}^{2} \tilde{L}_{i}^{I*} \tilde{L}_{i}^{I} - m_{R^{I}}^{2}
\tilde{R}^{I*} \tilde{R}^{I}  \nonumber  \\
 &  & -m_{Q^{I}}^{2} \tilde{Q}_{i}^{I*} \tilde{Q}_{i}^{I} - m_{D^{I}}^{2}
\tilde{D}
 ^{I*} \tilde{D}^{I} - m_{U^{I}}^{2}\tilde{U}^{I*} \tilde{U}^{I} + (m_{1}
\lambda_{B}
\lambda_{B}  \nonumber  \\
 &  & + m_{2}\lambda_{A}^{i}\lambda_{A}^{i} + m_{3} \lambda_{G}^{a}\lambda_{
 G}^{a} + h.c.) + \{ B\mu \varepsilon_{ij}H_{i}^{1}H_{j}^{2} + B_{I}\epsilon_{
I}\varepsilon_{ij}H_{i}^{2}\tilde{L}_{j}^{I}  \nonumber \\
 & & + \varepsilon_{ij} l_{sI}\mu H_{i}^{1}\tilde{L}_{j}^{I}\tilde{R}^{I} +
  d_{sI}\mu (-H_{1}^{1}\tilde{Q}_{2}^{I} + C^{IK}H_{2}^{1}\tilde{Q}_{1}^{K})
 \tilde{D}^{I}  \nonumber  \\
 & &+ u_{sI}\mu (-C^{KI*}H_{1}^{2}\tilde{Q}_{2}^{I} + H_{2}^{2}\tilde{Q}_{1}
 ^{I})
 \tilde{U}^{I} + h.c.\} . \label{softnew}
 \end{eqnarray}
 For simplicity we take from now on $\epsilon_{1} = \epsilon_{2} = 0$, in
 this way, that only $\tau$-lepton number is violated. The electroweak symmetry
 is broken when the two Higgs doublets $H^{1}$, $H^{2}$ and the $\tau$-slepton 
 acquire vacuum expectation values (VEVs)
\begin{equation}
H^{1}=
\left(
\begin{array}{c}
\frac{1}{\sqrt{2}}(\chi_{1}^{0} + \upsilon_{1} + i\varphi_{1}^{0})  \\
H_{2}^{1}
\end{array}
\right)
\end{equation}
\begin{equation}
H^{2}=
\left(
\begin{array}{c}
H^{2}_{1}  \\
\frac{1}{\sqrt{2}}(\chi_{2}^{0} + \upsilon_{2} + i\varphi_{2}^{0})
\end{array}
\right)
\end{equation}
\begin{equation}
\tilde{L}_{3}=
\left(
\begin{array}{c}
\frac{1}{\sqrt{2}}(\chi_{3}^{0} + \upsilon_{3} + i\varphi_{3}^{0})  \\
\tilde{\tau}_{L}^{-}
\end{array}
\right)
\end{equation}
Note that the gauge bosons $W$ and $Z_{0}$ acquire masses given by $m_{W}^{2}=
\frac{1}{4}g^{2}\upsilon^{2}$ and $m_{Z}^{2}=\frac{1}{4}(g^{2}+g^{\prime 2})
\upsilon^{2}$, where $ \upsilon^{2}=\upsilon_{1}^{2}+\upsilon_{2}^{2}
+\upsilon_{3}^{2}$ and $g$, $g^{\prime}$ are coupling constants of SU(2) and U(1).
We introduce the following notation in spherical coordinates\cite{s3}
\begin{eqnarray}
\upsilon_{1} & = & \upsilon \cos\theta_{\upsilon}\cos{\beta}, \nonumber  \\
\upsilon_{2} & = & \upsilon \sin{\beta} ,\nonumber  \\
\upsilon_{3} & = & \upsilon \sin\theta_{\upsilon}\cos\beta .
\end{eqnarray}
When the angle $\theta_{\upsilon}$ equals to zero, this sector will change back
to the MSSM limit exactly. The massless neutral Goldstone boson can be written as:
\begin{equation}
G^{0}=\cos\theta_{\upsilon}\cos\beta \varphi_{1}^{0} - 
\sin\beta \varphi_{2}^{0} + \sin\theta_{\upsilon}\cos\beta \varphi_{3}^{0} .
\end{equation}

In the model with bilinear R-parity violation, the charged Higgs bosons mix
with the left and right handed $\tau$-slepton.  In the original basis, where $\Phi_{c}=(H_
{2}^{1*}, H_{1}^{2}, \tilde{\tau}^{*}_{L}, \tilde{\tau}_{R})$, the scalar 
potential contains the following mass term
\begin{equation}
{\cal L}_{m}^{C} = -\Phi_{c}^{\dag}{\cal M}_{c}^{2}\Phi_{c}
\end{equation}
with the symmetric matrix ${\cal M}_{c}^{2}$ is given by 
(here the matrix is too big to be written in full so we write it by 
each element individually)
\begin{eqnarray}
{\cal M}_{c 1,1}^{2} & = & \frac{g^{2}}{4}(\upsilon_{2}^{2}-\upsilon_{3}^{2}) + 
\frac{1}{2}\upsilon_{3}^{2}l_{3}^
{2} +\epsilon_{3}\mu\frac{\upsilon_{3}}{\upsilon_{1}}+B\mu\frac{\upsilon_{2}}{
\upsilon_{1}}, \nonumber  \\
{\cal M}_{c 1,2}^{2} &=&\frac{g^{2}}{4}\upsilon_{1}\upsilon_{2} + B\mu , \nonumber \\
{\cal M}_{c 1,3}^{2} &=& \frac{g^{2}}{4}\upsilon_{1}\upsilon_{3} -\epsilon_{3}\mu-\frac{
1}{2}l_{3}\upsilon_{1}\upsilon_{3}  , \nonumber \\
{\cal M}_{c 1,4}^{2} &=& l_{3}\epsilon_{3}\frac{\upsilon_{2}}{\sqrt{2}} +
l_{s3}\frac{\mu \upsilon_{3}}{\sqrt{2}} , \nonumber \\
{\cal M}_{c 2,2}^{2} & = & \frac{g^{2}}{4}(\upsilon_{1}^{2}+\upsilon_{3}^{2}) - 
B_{3}\epsilon_{3}
\frac{\upsilon_{3}}{\upsilon_{2}} + B\mu\frac{\upsilon_{1}}{\upsilon_{2}}, \nonumber \\
{\cal M}_{c 2,3}^{2} &=& \frac{g^{2}}{4}\upsilon_{2}\upsilon_{3} - B_{3}\epsilon_{3} ,
\nonumber  \\
{\cal M}_{c 2,4}^{2} &=& \frac{l_{3}}{\sqrt{2}}\mu\upsilon_{3} + \frac{
l_{3}}{\sqrt{2}}\epsilon_{3}\upsilon_{1} ,\nonumber \\
{\cal M}_{c 3,3}^{2} & = & \frac{g^{2}}{4}(\upsilon_{2}^{2}-\upsilon_{1}^{2})+
\epsilon_{3}\frac{\mu\upsilon_{1}}{
\upsilon_{3}} - B_{3}\frac{\epsilon_{3}\upsilon_{2}}{\upsilon_{3}} + 
\frac{l_{3}^{2}}{2}\upsilon_{1}^{2}, \nonumber \\
{\cal M}_{c 3,4}^{2} &=& \frac{1}{\sqrt{2}}l_{3}\mu\upsilon_{2}-\frac{1}{
\sqrt{2}}l_{s3}\mu\upsilon_{1} , \nonumber  \\
{\cal M}_{c 4,4}^{2} &=& -\frac{g'^{2}}{4}(\upsilon_{1}^{2}-\upsilon_{2}^{2}+\upsilon
_{3}^{2})+\frac{1}{2}l_{3}^{2}(\upsilon_{1}^{2} + \upsilon_{3}^{2})
+ m_{R^{3}}^{2} .  
\label{eq-21}
\end{eqnarray}
This matrix has an eigenstate:
\begin{eqnarray}
G^{+} &=& \sum_{i=1}^{4} Z_{H}^{1,i} \Phi^{i}_{c}  \nonumber  \\
      &=& \frac{1}{\upsilon}(\upsilon_{1} H_{2}^{1*} - \upsilon_{2} H_{1}^{2}
         +\upsilon_{3}\tilde{\tau}_{L}^{*})  \nonumber  \\
      &=& \cos\theta_{\upsilon}\cos\beta H_{2}^{1*}-\sin\beta
       H_{1}^{2}+\sin\theta_{\upsilon}\cos\beta\tilde{\tau}_{L}^{*}   \label{goldc}
\end{eqnarray}
with zero eigenvalue, it is the massless charged Goldstone boson. In the physical
 (unitary) gauge, $G^{\pm}$ are absorbed by $W^{\pm}$ bosons and disappear from the
Lagrangian. The other three eigenstates $H^{+}$, $\tilde{\tau}_{1}$, 
$\tilde{\tau}_{2} $ can be expressed as:
\begin{eqnarray}
H^{+} &=& \sum_{i=1}^{4} Z_{H}^{2,i} \Phi^{i}_{c} ,\nonumber  \\
\tilde{\tau}_{1} &=& \sum_{i=1}^{4} Z_{H}^{3,i} \Phi^{i}_{c} ,\nonumber  \\
\tilde{\tau}_{2} &=& \sum_{i=1}^{4} Z_{H}^{4,i} \Phi^{i}_{c} . \label{chargh}
\end{eqnarray}
Similarly to the Higgs bosons, charginos mix with $\tau$ lepton forming a 
set of three charged fermions $\tau, \tilde{\kappa}_{2}^{-}, \tilde{\kappa}_{3}^{-}$ 
\cite{s13,s19}.
In the original basis where $\psi^{+T}=(-i\lambda^{+}$, $\tilde{H}_{2}^{1}$, $\tau_{
R}^{+})$ and $\psi^{-T}=(-i\lambda^{-}$, $\tilde{H}_{1}^{2}$, $\tau_{L}^{-})$, the 
charged fermion mass terms in the Lagrangian are
\begin{equation}
{\cal L}_{m}=-\psi^{-T}{\cal M}_{f} \psi^{+} + h.c.
\end{equation}
with the mass matrix is given by\cite{s13,s19}:
\begin{equation}
{\cal M}_{f} = 
\left(
\begin{array}{ccc}
2m_{2} & \frac{e\upsilon_{2}}{\sqrt{2}s_{W}} & 0  \\
\frac{e\upsilon_{1}}{\sqrt{2}s_{W}} & \mu & \frac{l_{3}\upsilon_{3}}{\sqrt
{2}}  \\
\frac{e\upsilon_{3}}{\sqrt{2}s_{W}} & \epsilon_{3} & \frac{l_{3}\upsilon_{1}
}{\sqrt{2}}  
\end{array}
\right)   \label{chaeig}
\end{equation}
where $S_{W}=\sin\theta_{W}$ and $\lambda^{\pm}=\frac{\lambda_{A}^{1} \mp i\lambda_{A}^{2}}{
\sqrt{2}}$. Also the two mixing matrices $Z^{+}$ and $ Z^{-} $ appear in the Lagrangian, 
they
are defined by the condition that the product $ (Z^{+})^{T}{\cal M}_{f} Z^{-}$
should be diagonal
\begin{equation}
(Z^{+})^{T}{\cal M}_{f} Z^{-} =
\left(
\begin{array}{ccc}
m_{\tilde{\kappa}_{1}^{-}} & 0 & 0 \\
0 & m_{\tilde{\kappa}_{2}^{-}} & 0 \\
0 & 0 & m_{\tilde{\kappa}_{3}^{-}}  
\end{array}
\right)  \label{zpm}
\end{equation}
Here, we assume $m_{\tilde{\kappa}_{1}^{-}} = m_{\tau}$ and $ m_{
\tilde{\kappa}_{3}^{-}} > m_{\tilde{\kappa}_{2}^{-}} > m_{\tilde{\kappa}_{1}^{-}} $. 

\section{Complete analysis of the rare B-processes in SUSY model with bilinear R-parity
violation}

Because we neglect the generation mixing of sleptons and squarks, there are no 
the couplings such as $\bar{\kappa}_{i}^{0}\gamma_{\mu} P_{L,R}q_{u}^{I}\tilde{U}_{1, 2}^{J}$ ($I \neq J$),  
$\bar{\kappa}_{i}^{0}\gamma_{\mu} P_{L,R}q_{d}^{I}\tilde{D}_{1, 2}^{J}$ ($I \neq J$),  
$\bar{\kappa}_{i}^{0}\gamma_{\mu}P_{L,R}e^{I}\tilde{E}_{1, 2}^{J}$ ($I \neq J$), 
$\bar{\lambda}_{G}^{a}\gamma_{\mu} P_{L,R}q_{u}^{I}\tilde{U}_{1, 2}^{J}$ ($I \neq J$) and 
$\bar{\lambda}_{G}^{a}\gamma_{\mu} P_{L,R}q_{d}^{I}\tilde{D}_{1, 2}^{J}$ 
($I \neq J$) where $I$, $J=1$, $2$, $3$ are the generational indices and $\kappa_{1}^{0}$, 
$\kappa_{2}^{0}$, $\kappa_{3}^{0}$, $\kappa_{4}^{0}$, $\lambda_{G}^{a}$ represent the
neutralinos and gluinos. Under this assumption, the contribution of gluino and neutralino 
is zero when we compute the rare processes such as
$b\rightarrow s+\gamma$, $b\rightarrow s+e^{+}e^{-} $ and $b \rightarrow s+\sum\nu_{i}\bar{\nu}_{i}$
at one loop level. 

In this section, we will use the effective Hamiltonian theory to discuss those processes. 
The method of the effective Hamiltonian theory was first used by Ref\cite{s18} and has
been developed over the last years\cite{s25}\cite{s26}. It is a two step program, starting with
an operator product expansion (OPE) and performing a renormalization group equation (RGE) analysis
afterwards. The derivation starts as follows: if the kinematics of the decay are of the kind 
that the masses of the internal particles $m_{i}$ are much larger than the external momentum
$p$: $m_{i}^{2} \gg p^{2}$, then the heavy particles can be integrated out. This concept
takes a concrete form with the functional integral formalism. It means that the heavy 
particles are removed as dynamical degrees of freedom from the theory, hence their fields
do not appear in the (effective) Lagrangian anymore. Their residual effect lies in the generated
effective vertices. In this way an effective low energy theory can be constructed from the 
full theory. In the framework of the standard model, strong interactions are known to
give sizable contributions to FCNC processes, the inclusion of the QCD corrections
increases the electroweak rate by about a factor two for $b\rightarrow s\gamma$
process, and enhances the rate about 30\% for $b\rightarrow se^{+}e^{-}$ 
transition\cite{s27}. Now, let us derive the 
effective Lagrangian for $b \rightarrow s\gamma$, $b \rightarrow se^{-}e^{+} $ and 
$b\rightarrow s\sum \nu_{i}\bar{\nu}_{i}$ first (at the $M_{W}$ scale),  then we analyze the 
renormalization group equation (RGE) in the leading logarithmic approximation (LLA), i.e. 
to sum up the terms $[\alpha_{s}\ln(\frac{M_{W}}{\mu})]^{n}$ to all orders n (n=0, 1, 
$\cdots$, $\infty$) in perturbation theory.

\subsection{The $b\bar{s} Z$, $b\bar{s} \gamma$ couplings and box diagrams in SUSY model with
bilinear R-parity violation}

\begin{figure}
\setlength{\unitlength}{1mm}
\begin{picture}(180,200)(0,0)
\put(-30,-40){\includegraphics{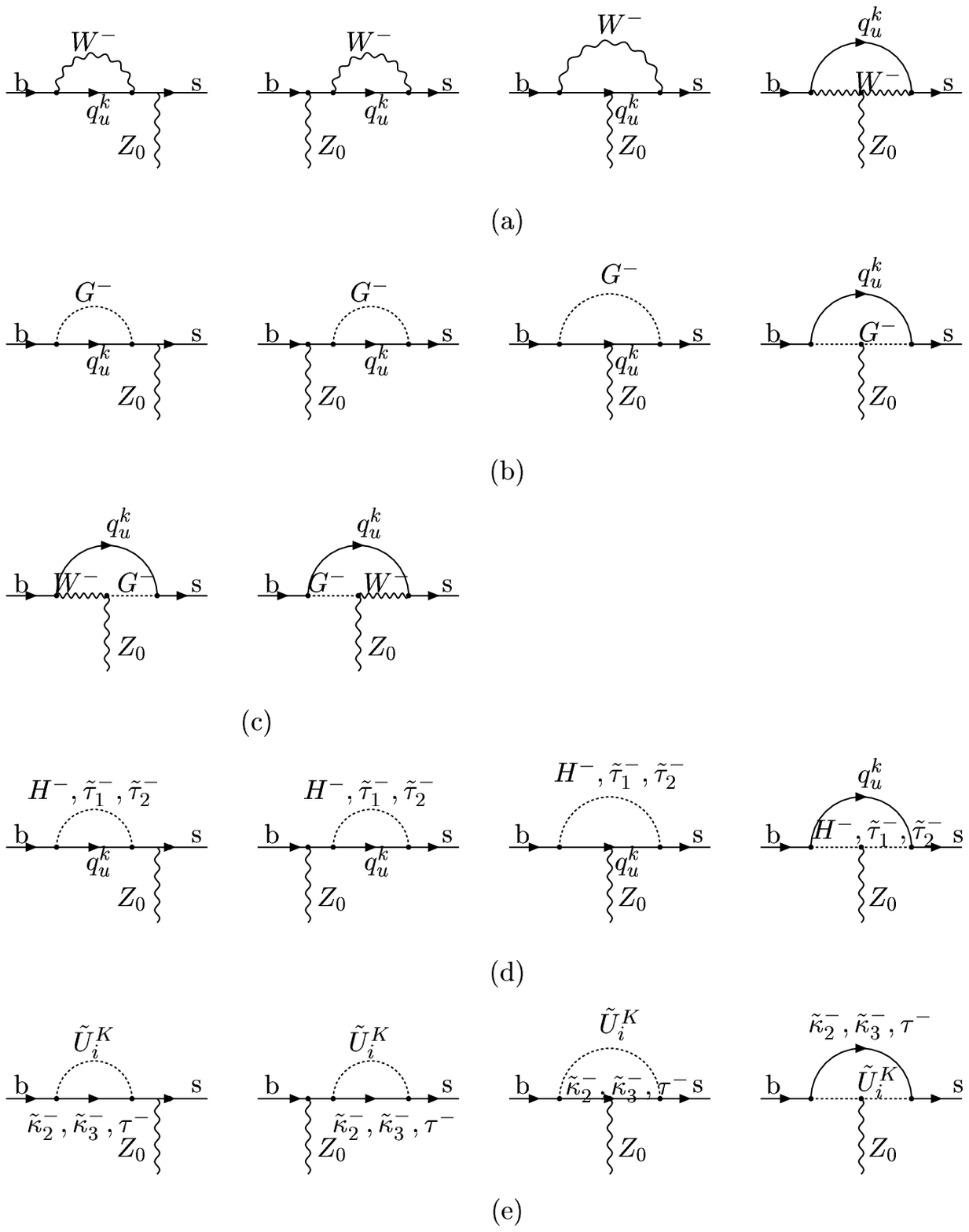}}
\end{picture}
\caption{The Feynman diagrams that contribute to $b\bar{s} \gamma$ and $b\bar{s} Z^{0}$ coupling}
\label{fig1}
\end{figure}

We follow the method that was used in Ref\cite{s18}. In the $'$t Hooft Feynman gauge,
the one-loop diagrams for the induced $b\bar{s} Z$ coupling are shown in Fig.\ \ref{fig1}. 
The diagrams Fig.\ \ref{fig1}(a), Fig.\ \ref{fig1}(b) and Fig.\ \ref{fig1}(c) represent
the SM contribution of the $b\rightarrow s$ transition. In each group, the first two
diagrams (except for Fig.\ \ref{fig1}(c)) are self-energy part. Since the weak current
is not conserved, we only need to expect a non-vanishing zeroth-order contribution in
the momentum q (q denotes the momentum transfer through the Z-boson). The induced 
$b\bar{s}Z$ coupling takes the form:
\begin{equation}
\Gamma^{(i)}_{Z_{\mu}}=\bar{s}\gamma_{\mu}P_{L}b\Gamma^{(i)}
\end{equation}
where $i=a$, $b$, $c$, $d$, $e$ and $P_{L,R} = \frac{1 \mp \gamma_{5}}{2}$. 
The $\Gamma^{i}$ can be written as (at the ${\rm M}_{W}$ scale):
\begin{eqnarray}
&\Gamma^{(a)} =& \frac{e^{3}}{(4\pi)^{2}\sin^{3}\theta_{W}\cos\theta_{W}}C_{ts}^{*}C_{tb}
\{(\frac{1}{2}-\frac{1}{3}\sin^{2}\theta_{W})[2f_{2}^{(0)}
(x_{tw}) -  \nonumber  \\
& & 4f_{3b}^{(1)}(x_{tw})]-(\frac{1}{4}-\frac{1}{3}
   \sin^{2}\theta_{W})f_{3b}^{(1)}(x_{tw})-  \nonumber  \\
& &\frac{2}{3}\sin^{2}\theta_{W}x_{tw}f_{3b}^{(0)}(x_{tw})+\frac{3}{2}\cos^{2}
   \theta_{W}f_{3a}^{(1)}(x_{tw})\}  ,\label{ga}  \\
& \Gamma^{(b)} =& \frac{e^{3}}{(4\pi)^{2}\sin^{3}\theta_{W}\cos\theta_{W}}C_{ts}^{*}C_{tb}
\{(\frac{1}{2}-\frac{1}{3}\sin^{2}\theta_{W}) \frac{x_{tw}}{2}[f_{2}^{(0)}(
x_{tw})-2f_{3b}^{(1)}(x_{tw})]+ \nonumber  \\
 & & \frac{x_{tw}}{2}[(\frac{1}{2}-\frac{2}{3}\sin^{2}\theta_{W})x_{tw}f_{3b}^{(0)}(
x_{tw})+\frac{1}{3}\sin^{2}\theta_{W}f_{3b}^{(1)}(x_{tw})
]+   \nonumber  \\
& &\frac{x_{tw}}{8}(\cos^{2}\theta_{W}-\sin^{2}\theta_{W})f_{3a}^{(1)}(x_{tw})\}  ,\label{gb} \\
& \Gamma^{(c)} =& \frac{e^{3}}{(4\pi)^{2}\sin^{3}\theta_{W}\cos\theta_{W}}C_{ts}^{*}C_{tb}
\sqrt{2}\sin^{2}\theta_{W}x_{tw}f_{3a}^{(0)}(x_{tw})  ,\label{gc}  \\
& \Gamma^{(d)} =& \frac{e^{3}}{(4\pi)^{2}\sin^{3}\theta_{W}\cos\theta_{W}}C_{ts}^{*}C_{tb}
\{(\frac{1}{2}-\frac{1}{3}\sin^{2}\theta_{W})\frac{x_{tw}}{2\sin^{2}\theta_{v}\sin^{2}\beta}
\sum_{i=2}^{4}Z_{H}^{2i}Z_{H}^{2i*}[f_{2}^{(0)}(x_{ts_{i}})-  \nonumber \\
& &2f_{3b}^{(1)}(x_{ts_{i}})] + \frac{x_{tw}}{2\sin^{2}\theta_{v}
\sin^{2}\beta}\sum_{i=2}^{4}Z_{H}^{2i}Z_{H}^{2i*}[(\frac{1}{2}- 
\frac{2}{3}\sin^{2}\theta_{W})x_{ts_{i}}f_{3b}^{(0)}(x_{ts_{i}})+ \nonumber \\
& &\frac{1}{3}\sin^{2}\theta_{W}f_{3b}^{(0)}(x_{ts_{i}})]
- \frac{x_{tw}}{8\sin\theta_{v}\sin\beta} 
\sum_{i=2}^{4}\sum_{j=2}^{4}Z_{H}^{2i}Z_{H}^{2j*}[(\cos^{2}\theta
_{W}- \nonumber  \\
& &\sin^{2}\theta_{W})\delta_{ij}-Z_{H}^{4i}Z_{H}^{4j*}]f_{3c}^{(1)}(
x_{ts_{i}},x_{s_{j}s_{i}})\}  ,\label{gd} \\
& \Gamma^{(e)} =& \frac{e^{3}}{(4\pi)^{2}\sin^{3}\theta_{W}\cos\theta_{W}}C_{ts}^{*}C_{tb}
\{\sum_{i=1}^{2}\sum_{j=1}^{3}|(-Z_{\tilde{Q}^{3}}^{1i*}Z_{1j}^{+} + 
\frac{Z_{\tilde{Q}^{3}}^{2i*}Z^{+}_{2j}}{\sqrt{2}M_{W}\sin\theta_{v}
\sin\beta})|^{2}(\frac{1}{2}-  \nonumber \\ 
& &\frac{1}{3}\sin^{2}\theta_{W})[f_{2}^{(0)}(x_{\tilde{\kappa}^{-}_{j}\tilde{t}_{i}})- 
2f_{3b}^{(1)}(x_{\tilde{\kappa}_{j}^{-}\tilde{t}_{i}})] + 
\sum_{i,l=1}^{2}\sum_{j=1}^{3}(-Z_{\tilde{Q}^{3}}^{1i*}Z_{1j}^{+}+\frac{m_{t}
Z_{\tilde{Q}^{3}}^{2i*}Z_{2j}^{+}}{\sqrt{2}M_{W}\sin\theta_{v}\sin\beta}) \nonumber \\
& &(-Z_{\tilde{Q}^{3}}^{1l}Z_{1j}^{+*}+ 
\frac{m_{t}Z_{\tilde{Q}^{3}}^{2l}Z_{2j}^{+*}}{\sqrt{2}M_{W}\sin\theta_{v}\sin\beta})(-\frac{1}{4}Z_{\tilde{Q}^{3}}
^{1l*}Z^{1i}_{\tilde{Q}^{3}}-\frac{1}{3}\sin^{2}\theta_{W}\delta_{il})f_{3c}^{(0)}(x_{\tilde{\kappa}^{-}
_{j}\tilde{t}_{i}},x_{\tilde{t}_{l}\tilde{t}_{i}}) + \nonumber \\
& &\sum_{i,j=1}^{3}\sum_{l=1}^{2}\frac{1}{2}(-Z_{\tilde{Q}^{3}}^{1l}Z_{1j}^{+*}+
\frac{m_{t}Z_{\tilde{Q}^{3}}^{2l}Z_{2j}^{+*}}{\sqrt{2}M_{W}\sin\theta_{v}\sin\beta})(-Z_{\tilde{Q}^{3}}^{1l*}
Z_{1j}^{+}+\frac{m_{t}Z_{\tilde{Q}^{3}}^{2l*}Z_{2j}^{+}}{\sqrt{2}M_{W}\sin\theta_{v}\sin\beta}) \nonumber \\
& &[(Z_{1j}^{-}Z_{1i}^{-*}+2\delta_{ij}\cos2\theta_{W})\frac{1}{2}
f_{3c}^{(1)}(x_{\tilde{\kappa}_{i}^{-}\tilde{t}_{l}},x_{\tilde{\kappa}_{j}^{-}\tilde{
t}_{l}})+x_{\tilde{\kappa}_{i}^{-}\tilde{t}_{l}}x_{\tilde{\kappa}_{j}^{-}\tilde{t}_{l
}}(Z_{1j}^{+*}Z_{1i}^{+}  \nonumber \\
& &+2\delta_{ij}\cos2\theta_{W})f_{3c}^{(0)}(x_{\tilde{\kappa}_{i}
^{-}\tilde{t}_{l}},x_{\tilde{\kappa}_{j}^{-}\tilde{t}_{l}})]\}  ,\label{ge}
\end{eqnarray}
where $s_{i=1,2,3,4}=(G^{-}$, $H^{-}$, $\tilde{\tau}_{1}$, $\tilde{\tau}_{2})$ and $ x_{\alpha\beta}=\frac{m_{\alpha}^{2}}
{m_{\beta}^{2}}$. $Z_{\tilde{Q}^{3}}$ is the top-squark mixing matrix, its definition can be found
in appendix. B.

Notice that the ultraviolet divergences cancel separately for each of those equation.
In obtaining the form shown, the unitary property of $Z_{H}$, $Z_{\tilde{Q}^{3}}$, $Z^{+}$ and $Z^{-}$
has been used, together with the unitarity of the Kobayashi-Maskawa matrix.

The computation of the photon exchanged contribution is somewhat more involved and 
requires the calculation of the induced $b\bar{s}\gamma$ coupling up to second order 
in the external momentum. The diagrams need to be computed are those of Fig.\ \ref{fig1}
with $Z$ being replaced by $\gamma$. The induced $b\bar{s}\gamma$ coupling takes the
form (at the ${\rm M}_{W}$ scale):
\begin{equation}
\Gamma_{\gamma_{\mu}}^{(i)}=\bar{s}[F_{1}^{(i)}(q^{2}\gamma_{\mu}-q_{\mu}/\!\!\!{q})P_{L}
+ F_{2}^{(i)}/\!\!\!{q}\gamma_{\mu}(m_{s}P_{L}+m_{b}P_{R})]b
\end{equation}
where $i=a$, $b$, $c$, $d$, $e$ and 
\begin{eqnarray}
& F_{1}^{(a)}= & \frac{1}{(4\pi)^{2}}\frac{e^{3}}{\sin^{2}\theta_{W}}C_{tb}C_{ts}^{*}
\frac{1}{M_{W}^{2}}\{\frac{1}{9}f_{5d}^{(2)}(x_{tw})-\frac{2}{3}f_{4c}^{(1)}(x_{tw})+\nonumber \\
& &\frac{1}{2}f_{3a}^{(0)}(x_{tw}) + 
\frac{1}{2}f_{4b}^{(1)}(x_{tw})-\frac{3}{2}f_{4a}^{(1)}(x_{tw})+ \nonumber \\
& &\frac{4}{3}f_{5a}^{(2)}(x_{tw})-\frac{1}{3}f_{5b}^{(2)}(x_{tw})\}  ,\label{f1a}  \\
& F_{2}^{(a)}= & \frac{1}{(4\pi)^{2}}\frac{e^{3}}{\sin^{2}\theta_{W}}C_{tb}C_{ts}^{*}
\frac{1}{M_{W}^{2}}\{\frac{2}{3}f_{3b}^{(0)}(x_{tw})-f_{4c}^{(1)}(x_{tw})+ \nonumber  \\
& &\frac{1}{3}f_{5d}^{(2)}(x_{tw})+\frac{1}{2}f_{3a}^{(0)}(x_{tw}) - 
\frac{3}{2}f_{4b}^{(1)}(x_{tw})- \nonumber \\ 
& &\frac{1}{4}f_{4a}^{(1)}(x_{tw})+\frac{1}{3}
f_{5c}^{(2)}(x_{tw})+\frac{1}{3}f_{5b}^{(2)}(x_{tw})+\frac{2}{9}[\frac{\ln x_{tw}}{x_{tw}-1} \nonumber  \\
& &+ \ln \frac{m_{c}^{2}}{M_{W}^{2}} + f(\frac{q^{2}}{m_{b}^{2}})]\}  ,\label{f2a}  \\
& F_{1}^{(b)} = & \frac{1}{(4\pi)^{2}}\frac{e^{3}}{\sin^{2}\theta_{W}}C_{tb}C_{ts}^{*}
\frac{1}{M_{W}^{2}}\{-\frac{x_{tw}}{18}f_{5d}^{(2)}(x_{tw}) -
\frac{x_{tw}}{12}f_{5c}^{(2)}(x_{tw}) +  \nonumber  \\
& & \frac{x_{tw}}{12}f_{5b}^{(2)}(
x_{tw})\}  ,\label{f1b} \\
& F_{2}^{(b)} = & \frac{1}{(4\pi)^{2}}\frac{e^{3}}{\sin^{2}\theta_{W}}C_{tb}C_{ts}^{*}
\frac{1}{M_{W}^{2}}\{\frac{x_{tw}}{2}f_{4c}^{(1)}(x_{tw})-\frac{x_{tw}}{
6}f_{5d}^{(2)}(x_{tw})+  \nonumber  \\
& &\frac{3x_{tw}}{8}f_{4b}^{(1)}(x_{tw}) -  \frac{1}{6}f_{5c}^{(2)}(x_{tw})-  \nonumber  \\
& &\frac{1}{12}f_{5b}^{(2)}(x_{tw})\}  ,\label{f2b}  \\
& F_{1}^{(c)} = & -\frac{1}{(4\pi)^{2}}\frac{e^{3}}{\sin^{2}\theta_{W}}C_{tb}C_{ts}^{*}
\frac{x_{tw}}{M_{W}^{2}}f_{5a}^{(2)}(x_{tw})  ,\label{f1c}  \\
& F_{2}^{(c)}= & \frac{1}{(4\pi)^{2}}\frac{e^{3}}{\sin^{2}\theta_{W}}C_{tb}C_{ts}^{*}
\frac{1}{4M_{W}^{2}}f_{4a}^{(1)}(x_{tw})   ,\label{f2c} \\
& F_{1}^{(d)} = & \frac{1}{(4\pi)^{2}}\frac{e^{3}}{\sin^{2}\theta_{W}}C_{tb}C_{ts}^{*}
\sum_{i=2}^{4}\frac{Z_{H}^{2i}Z_{H}^{2i*}}{6M_{W}^{2}\sin^{2}\theta_{W}\sin^{2}\beta}
\{-\frac{1}{3}f_{5d}^{(2)}(x_{ts_{i}}) - \nonumber \\
& &\frac{1}{2}f_{5c}^{(2)}(x_{ts_{i}})+\frac{1}{2}f_{5b}^{(2)}(x_{ts_{i}})\}  ,\label{f1d} \\
& F_{2}^{(d)}= & \frac{1}{(4\pi)^{2}}\frac{e^{3}}{\sin^{2}\theta_{W}}C_{tb}C_{ts}^{*}
\sum_{i=2}^{4}\{\frac{x_{ts_{i}}}{6M_{W}^{2}\sin^{2}\theta_{W}}[
(\frac{Z_{H}^{2i}Z_{H}^{2i*}}{\sin^{2}\beta}+2\frac{Z_{H}^{1i*}
Z_{H}^{2i}}{\sin\beta \cos\beta})f_{4c}^{(1)}(x_{ts_{i}})-  \nonumber \\
& &\frac{Z_{H}^{2i}Z_{H}^{2i*}}{\sin^{2}\beta}f_{5d}^{(2)}(x_{ts_{i}})] + 
\frac{x_{ts_{i}}}{2M_{W}^{2}\sin^{2}\theta_{v}}[\frac{Z_{H}^{2i}Z_{H}^{2i*}}{4\sin^{2}
\beta}f_{4b}^{(1)}(x_{ts_{i}})+  \nonumber  \\
& &\frac{Z_{H}^{1i*}Z_{H}^{2i}}{2\sin\beta \cos
\beta}f_{4b}^{(1)}(x_{ts_{i}})-\frac{Z_{H}^{2i}Z_{H}^{2i*}}{3\sin^{2}\beta}
f_{5c}^{(2)}(x_{ts_{i}})- \nonumber  \\
& &\frac{Z_{H}^{2i}Z_{H}^{2i*}}{6\sin^{2}\beta}f_{5b}^{(2)}(x_{ts_{i}})]\}  ,\label{f2d} \\
& F_{1}^{(e)}= & \frac{1}{(4\pi)^{2}}\frac{e^{3}}{\sin^{2}\theta_{W}}C_{tb}C_{ts}^{*}
\sum_{i=1,2}\sum_{j=1}^{3}\frac{1}{m^{2}_{\tilde{t}_{i}}}|Z_{\tilde{Q}^{3}}^{1i}Z^{+*}_{1,j}-\frac{m_{t}Z_{\tilde{Q}^{3}}^{2i}
Z^{+*}_{2j}}{\sqrt{2}M_{W}\sin\theta_{v}\sin\beta}|^{2}\{\frac{2}{9}f_{5a}^{(2)}(x_{\tilde{
\kappa}_{j}^{-}\tilde{t}_{i}}) + \nonumber  \\
& &\frac{1}{6}f_{5d}^{(2)}(x_{\tilde{\kappa}_{j}^{-}\tilde{t}_{i}})\}  ,\label{f1e} \\
& F_{2}^{(e)}= & \frac{1}{(4\pi)^{2}}\frac{e^{3}}{\sin^{2}\theta_{W}}C_{tb}C_{ts}^{*}
\sum_{i=1,2}\sum_{j=1}^{3}\frac{1}{m_{\tilde{t}_{i}}^{2}}\{|Z_{\tilde{Q}^{3}}^{1i}Z^{+*}_{1,j}-\frac{m_{t}Z_{\tilde{Q}^{3}}^{2i}
Z^{+*}_{2j}}{\sqrt{2}M_{W}\sin\theta_{v}\sin\beta}|^{2}[-\frac{1}{2}f_{4c}^{(1)}(
x_{\tilde{\kappa}_{j}^{-}\tilde{t}_{i}})+  \nonumber  \\
& &\frac{1}{2}f_{5d}^{(2)}(x_{\tilde{\kappa}_{j}^{-}
\tilde{t}_{i}}) - 
\frac{2}{3}f_{4b}^{(1)}(x_{\tilde{\kappa}_{j}^{-}\tilde{t}_{i}}
)+\frac{2}{9}f_{5c}^{(2)}(x_{\tilde{\kappa}_{j}\tilde{t}_{i}}^{-})- \nonumber  \\
& &\frac{1}{18}f_{5b}^{(2)}(x_{\tilde{\kappa}_{j}^{-}\tilde{t}_{i}}
)]-\frac{1}{2}(Z_{\tilde{Q}^{3}}^{1i}Z_{1j}^{+*}+
\frac{m_{t}Z_{\tilde{Q}^{3}}^{2i}Z_{2j}^{+*}}{\sqrt{2}M_{W}\sin\theta_{v} \sin\beta})\frac{
Z_{\tilde{Q}^{3}}^{1i*}Z_{2j}^{-*}m_{\tilde{\kappa}_{j}}}{\sqrt{2}M_{W}\sin\theta_{v}\cos\beta}  \nonumber  \\
& &[f_{4c}^{(1)}(x_{\tilde{\kappa}_{j}^{-}\tilde{t}_{i}}) + 
\frac{2}{3}f_{4b}^{(1)}(x_{\tilde{\kappa}_{j}^{-}\tilde{t}_{i}})]\} . \label{f2e}
\end{eqnarray}

The function f(s) is defined as
\begin{equation}
f(s) = -\frac{2}{3} -\frac{z}{s} + 
\left\{ 
\begin{array}{lll}
2(1+\frac{z}{2s})\sqrt{\frac{z}{s}-1} \tan^{-1} [ \sqrt{ \frac{z}{s}-1} ]^{-1}, & {\rm if} & s < z , \\
(1+\frac{z}{2s})\sqrt{1-\frac{z}{s}} [ \ln \frac{1+\sqrt{1-z/s}}{1-\sqrt{1-z/s}} -i\pi ], & {\rm if} &  s > z 
\end{array} 
\right.     
\end{equation}
where $z=\frac{4m_{c}^{2}}{m_{b}^{2}} $ and $f(0)=1$. The other functions are given in appendix. A.

\begin{figure}
\setlength{\unitlength}{1mm}
\begin{picture}(180,200)(0,0)
\put(-30,-40){\includegraphics{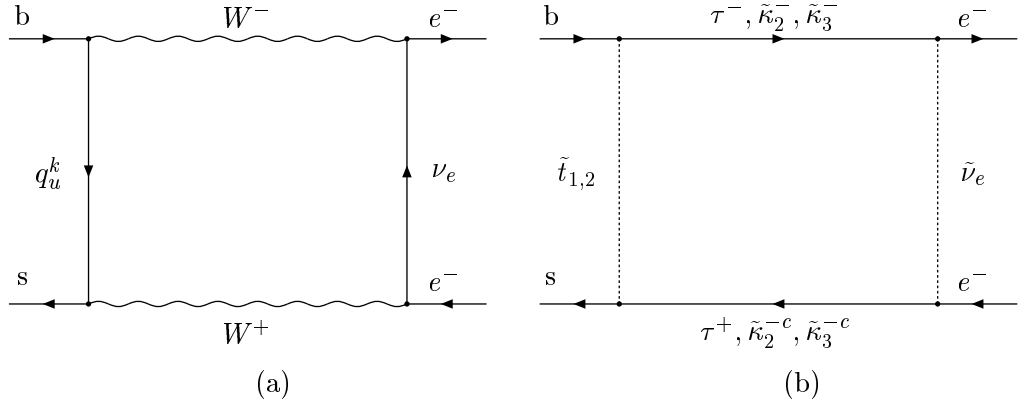}}
\end{picture}
\caption{The box diagrams contribute to $b\rightarrow se^{+}e^{-}$}
\label{fig2}
\end{figure}

The box-diagrams that contribute to the $b\rightarrow se^{+}e^{-}$ are shown in 
Fig.\ \ref{fig2}. The effective Lagrangian takes the form (at the ${\rm M}_{W}$ scale):
\begin{equation}
A_{i}^{box}(\bar{s}\gamma^{\mu}P_{L}b)(\bar{e}\gamma_{\mu}P_{L}e)
\end{equation}
with
\begin{eqnarray}
&A_{a}^{box} =& \frac{1}{(4\pi)^{2}}\frac{e^{4}}{4\sin^{4}\theta_{W}}C_{ts}^{*}C_{tb}
\frac{1}{M_{W}^{2}}f_{4d}^{(1)}(x_{tw},0)  ,\label{boxea} \\
& A_{b}^{box} =& \frac{1}{(4\pi)^{2}}\frac{e^{4}}{4\sin^{4}\theta_{W}}C_{ts}^{*}C_{tb}
\sum_{i,j=1}^{3}\sum_{l=1}^{2}\frac{1}{m_{\tilde{t}_{l}}^{2}}Z_{1j}^{+*}Z_{1i}^{+}
(-Z_{\tilde{Q}^{3}}^{1l*}Z^{+}_{1j}+\frac{m_{t}Z_{\tilde{Q}^{3}}^{2l*}Z_{2i}^{+}}{\sqrt{2}M_{W}\sin\theta_{v}
\sin\beta})  \nonumber  \\
& &(-Z_{\tilde{Q}^{3}}^{1l}Z^{+*}_{1j}+\frac{m_{t}Z_{\tilde{Q}^{3}}^{2l}Z_{2i}^{+*}}{\sqrt{2}M_{W}\sin\theta_{v}
\sin\beta})f_{4e}^{(1)}(x_{\tilde{\kappa}_{j}^{-}\tilde{t}_{l}},x_{\tilde{\kappa}_{i}^{-}\tilde{t}_{l}}
,x_{\tilde{\nu}_{e}\tilde{t}_{l}})  .\label{boxeb}
\end{eqnarray}

\begin{figure}
\setlength{\unitlength}{1mm}
\begin{picture}(180,200)(0,0)
\put(-30,-40){\includegraphics{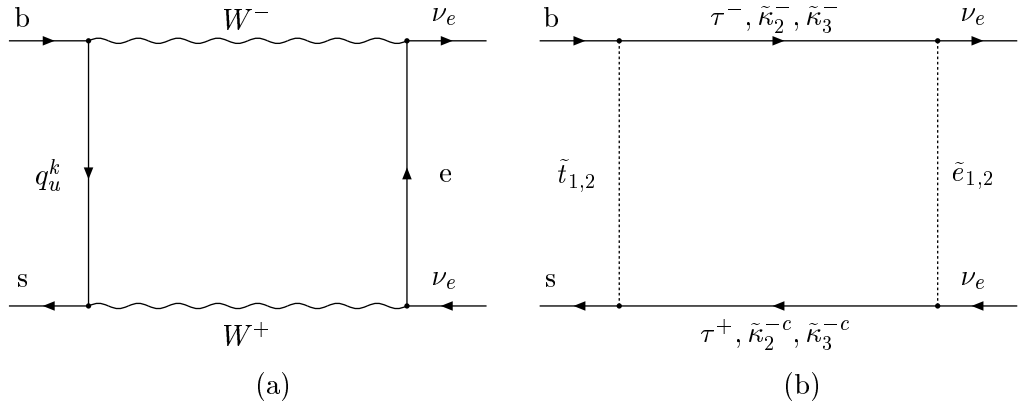}}
\end{picture}
\caption{The box diagrams contribute to $b\rightarrow s\nu_{e} \bar{\nu}_{e}$}
\label{fig3}
\end{figure}

As for $b\rightarrow s\nu_{e}\bar{\nu}_{e}$, it is analogous with the case of $b\rightarrow
se^{+}e^{-}$. The box diagrams that contribute to the $b\rightarrow s\nu_{e}\bar{\nu}_{e}$
are given in Fig.\ \ref{fig3}. The effective Lagrangian takes the form (at the ${\rm M}_{W}$ scale):
\begin{equation}
B_{i}^{box}(\bar{s}\gamma^{\mu}P_{L}b)(\bar{\nu}_{e}\gamma_{\mu}P_{L}\nu_{e})
\end{equation}
with
\begin{eqnarray}
& B_{a}^{box} =& \frac{1}{(4\pi)^{2}}\frac{e^{4}}{4\sin^{4}\theta_{W}}C_{ts}^{*}C_{tb}
\frac{1}{M_{W}^{2}}f_{4d}^{(1)}(x_{eW},x_{tw}) , \label{boxna} \\
& B_{b}^{box} =&\frac{1}{(4\pi)^{2}}\frac{e^{4}}{4\sin^{4}\theta_{W}}C_{ts}^{*}C_{tb}
\frac{1}{m_{\tilde{t}_{l}}^{2}}\sum_{h=1,2}\sum_{i,j}^{3}\sum_{l=1,2}Z_{\tilde{E}^{1}}^{1h}Z_{\tilde{E}^{1}}
^{1h*}Z_{1i}^{-}Z_{1j}^{-*}  \nonumber  \\
& &|-Z_{\tilde{Q}^{3}}^{1l*}Z_{1i}^{+}+\frac{m_{t}Z_{\tilde{Q}^{3}}^{2l*}
Z_{2i}^{+}}{\sqrt{2}M_{W}\sin\theta_{v}\sin\beta}|^{2}f_{4e}^{(1)}(x_{\tilde{\kappa}_{i}^{-}
\tilde{t}_{l}},x_{\tilde{\kappa}_{j}^{-}\tilde{t}_{l}},x_{\tilde{e}_{h}^{-}
\tilde{t}_{l}})   \label{boxnb}
\end{eqnarray}
where $Z_{\tilde{E}^{I}}$ is the mixing matrix of slepton (I=1, 2 is the index of generation), 
 its definition can be found in appendix. B. 

\subsection{The width of $b\rightarrow s+\gamma$ in SUSY model with bilinear R-parity violation}

The total amplitude of the decay $b\rightarrow s+\gamma$ can therefore be written as:
\begin{equation}
{\cal A}_{tot}^{\gamma}(b\rightarrow s+\gamma)=F_{2}{\cal O}_{LR}^{\gamma}
\end{equation}
with $F_{2}$ is the sum of Eq.\ (\ref{f2a}), Eq.\ (\ref{f2b}), Eq.\ (\ref{f2c}), Eq.\ (\ref{f2d}),
Eq.\ (\ref{f2e}). Where ${\cal O}_{LR}^{\gamma}=m_{b}\epsilon_{\mu}
\bar{s}/\!\!\!{q}\gamma^{\mu}P_{R}b$ and the contribution of ${\cal O}_{RL}^{\gamma}=m_{s}
\epsilon_{\mu}\bar{s}/\!\!\!{q}\gamma^{\mu}P_{L}b$ is neglected since it is order of $O(
\frac{m_{s}^{2}}{m_{b}^{2}})$.

By denoting the total amplitude at a scale $\mu$ as $F_{2}(\mu)$, the QCD- corrected
amplitude at the scale of the process ($\sim m_{b}$) is then given by\cite{s15}
\begin{equation}
F_{2}(m_{b})=\eta^{-\frac{16}{23}}\{F_{2}(M_{W})+F_{2}^{0}[\frac{116}{135}(\eta^{\frac{
10}{23}}-1)+\frac{58}{189}(\eta^{\frac{28}{23}}-1)]\} ,
\end{equation}
where 
\begin{eqnarray}
\eta &=& \alpha_{s}(m_{b})/\alpha_{s}(M_{W}) \approx 1.8 ,  \nonumber \\
F_{2}^{0}&=&\frac{1}{(4\pi)^{2}}\frac{e^{3}}{2\sin^{3}\theta_{W}}C_{tb}C_{ts}^{*}\frac{1}{
M_{W}^{2}}  .\label{etaf20}
\end{eqnarray}

The property $C_{cs}^{*}C_{cb} \approx -C_{ts}^{*}C_{tb}$ for the $3\times 3$ CKM matrix
has been used in the previous equation. The inclusive width for $b\rightarrow s +
\gamma $ decay is finally given by
\begin{equation}
\Gamma(b\rightarrow s+\gamma)=\frac{m_{b}^{5}}{16\pi}|F_{2}(m_{b})|^{2}
\end{equation}
Where we have neglected the phase-space factor of order $O(\frac{m_{s}^{2}}{m_{b}^{2}})$.
We calculate the corresponding branching ratio as in Ref\cite{s19} by making use of the 
semileptonic decay $b\rightarrow ce\bar{\nu}$, one gets:
\begin{equation}
Br(b\rightarrow s+\gamma)=\frac{\Gamma(b\rightarrow s+\gamma)}{\Gamma(b\rightarrow ce\bar{
\nu})}Br(b\rightarrow ce\bar{\nu})
\end{equation}
where for $Br(b\rightarrow ce\bar{\nu})$ we use the averaged experimental value 0.11
\cite{s20}. The QCD- corrected width for the semileptonic decay $b\rightarrow ce\bar{\nu}
$ is \cite{s21}
\begin{equation}
\Gamma(b\rightarrow ce\bar{\nu})=\frac{G_{F}^{2}m_{b}^{5}}{192 \pi^{3}}\rho(\frac{m_{c}}{
m_{b}},0,0)|C_{bc}|^{2}\{1-\frac{2\alpha_{s}(m_{b})}{3\pi}f(\frac{m_{c}}{m_{b}},0,0)\}
\end{equation}
where the phase-space factor $\rho$ is 0.447 and $f(\frac{m_{c}}{m_{b}},0,0)=2.41$, $G_{F}$
is the Fermi constant.

\subsection{The width of $b\rightarrow se^{+}e^{-}$ in SUSY model with bilinear R-parity
violation}

This decay has been often considered the benchmark of charmless b-decays with strange 
particles in the final state. We will provide below the amplitude for $b\rightarrow s
e^{+}e^{-}$ including QCD effects. In what follows, the same conventions as in the 
previous decay have been adopted.

We begin by considering the diagrams which induce the effective flavor-changing coupling
of the photon to quarks (photon penguins). They are given by the diagrams shown in
Fig.\ \ref{fig1} with a lepton line attached to the photon propagator. We consider separately 
the monopole (LL) and dipole (LR) form factor, which are related to different effective
operators.\\
(i)Photon penguins (LL component, at the ${\rm M}_{W}$ scale)
\begin{eqnarray}
{\cal A}_{tot}^{\gamma,LL}(b\rightarrow se^{+}e^{-})& = &eF_{1}{\cal O}_{LLV}  \nonumber \\
 & \equiv &F_{1}^{\gamma}{\cal O}_{LLV}
\end{eqnarray}
where ${\cal O}_{LLV}=(\bar{s}\gamma^{\mu}P_{L}b)(\bar{e}\gamma_{\mu}e)$ and $F_{1}$ is the sum of Eq.\ (\ref{f1a}),
Eq.\ (\ref{f1b}), Eq.\ (\ref{f1c}), Eq.\ (\ref{f1d}), Eq.\ (\ref{f1e}). \\
(ii)Photon penguins (LR component, at the ${\rm M}_{W}$ scale)
\begin{eqnarray}
{\cal A}_{tot}^{\gamma,LR}(b\rightarrow se^{+}e^{-})& = &eF_{2}{\cal O}_{LRV}  \nonumber  \\
 & \equiv & F_{2}^{\gamma}{\cal O}_{LRV}
\end{eqnarray}
 with $F_{2}$ is the sum of Eq.\ (\ref{f2a}), Eq.\ (\ref{f2b}), Eq.\ (\ref{f2c}), Eq.\ (\ref{f2d}),
 Eq.\ (\ref{f2e}) and the operator ${\cal O}_{LRV}$ is defined as: 
\[ {\cal O}_{LRV} \equiv m_{b}\frac{1}{q^{2}}(\bar{s}/\!\!\!{q}\gamma^{\mu}P_{R}b)(\bar{e}
\gamma_{\mu}e)  \]
(iii)$Z^{0}$- penguins. The process $b\rightarrow se^{+}e^{-}$ is also induced by the 
effective FC coupling of the $Z^{0}$ to quarks. The total amplitude coming from the $Z^{0}$
-penguins can be expressed as (at the ${\rm M}_{W}$ scale)
\begin{eqnarray}
{\cal A}_{tot}^{Z}(b\rightarrow se^{+}e^{-}) & = & \frac{e}{M_{Z}^{2} \sin\theta_{W}\cos\theta_{W}}
\Gamma_{Z}(-\frac{1}{2}{\cal O}_{LLL}+\sin^{2}\theta_{W}{\cal O}_{LLV})  \nonumber  \\
 & \equiv & A^{Z}(-\frac{1}{2}{\cal O}_{LLL}+\sin^{2}\theta_{W}{\cal O}_{LLV})
\end{eqnarray}
where the new operator ${\cal O}_{LLL}$ is given as \[ {\cal O}_{LLL}=(\bar{s}\gamma^{\mu}P_{L}b)(
\bar{e}\gamma_{\mu}P_{L}e) \] and $\Gamma_{Z}=\sum \Gamma_{Z}^{(i)} $ is the sum of 
Eq.\ (\ref{ga}), Eq.\ (\ref{gb}),Eq.\ (\ref{gc}), Eq.\ (\ref{gd}) and Eq.\ (\ref{ge}). \\
(iv)Box diagrams. The relevant contributions are given in Eq.\ (\ref{boxea}),
Eq.\ (\ref{boxeb}).  \\

In order to implement the QCD corrections, let us rewrite the total amplitude at the $M_{
W}$ scale as:
\begin{equation}
{\cal A}_{tot}^{(b\rightarrow se^{+}e^{-})}(M_{W})=A_{LLV}(M_{W}){\cal O}_{LLV} + A_{LRV}(M_{W})
{\cal O}_{LRV} + A_{LLL}(M_{W}){\cal O}_{LLL}
\end{equation}
with
\begin{eqnarray}
A_{LLV}(M_{W}) &=& F_{1}^{\gamma} + \sin^{2}\theta_{W}A^{Z} ,\nonumber \\
A_{LRV}(M_{W}) &=& F_{2}^{\gamma}  ,\nonumber \\
A_{LLL}(M_{W}) &=& -A^{Z}/2+(A^{box}_{a}+A^{box}_{b}) .
\end{eqnarray}
Renormalization at the $m_{b}$-scale leads to\cite{s22}
\begin{eqnarray}
&A_{LLV}(m_{b}) =& A_{LLV}^{\gamma}(M_{W}) + \sin^{2}\theta_{W}A^{Z}(M_{W}) + A_{0}'\frac{
4\pi}{\alpha_{s}(M_{W})}\{\frac{8}{87}[1-\eta^{-\frac{29}{23}}] \nonumber \\
& &-\frac{4}{33}[1-\eta^{-\frac{11}{23}}]\}-\frac{4}{9}A_{0}'[\ln (\frac{m_{c}^{2}}{m_{b}^{2}})+
f(s)][2\eta^{-\frac{6}{23}} - \nonumber  \\
& &\eta^{\frac{12}{23}} ] , \\
& A_{LRV}(m_{b})= & \eta^{-\frac{16}{23}}\{A_{lRV}(M_{W})+A_{0}'[\frac{116}{135}(\eta^{\frac{10}{
23}}-1)+\frac{58}{189}(\eta^{\frac{28}{23}}-1)]\} , \\
& A_{LLL}(m_{b})= & A_{LLL}(M_{W}) ,
\end{eqnarray}
where $A_{0}'=F_{2}^{0}\sqrt{4\pi\alpha}$ and $\eta, F_{2}^{0}$ have been defined
in Eq.\ (\ref{etaf20}) .

The differential decay rate is then given by\cite{s15}:
\begin{eqnarray}
\frac{d\Gamma}{ds} &=& \frac{m_{b}^{5}}{1536\pi^{3}}4(1-s^{2})\{(\frac{1}{2}+s)[|A_{LLV}
+A_{LLL}|^{2}+|A_{LLV}|^{2}]  \nonumber  \\
& & +(1+\frac{2}{s})|A_{LRV}|^{2}-3Re[(2A_{LLV}+A_{LLL})A_{LRV}^{*}]\}
\end{eqnarray}
where $s=\frac{q^{2}}{m_{b}^{2}}$.

\subsection{The width of $b\rightarrow s\sum \nu_{i}\bar{\nu}_{i}$ in SUSY model with bilinear 
R-parity violation}

The transition $b\rightarrow s\sum \nu_{i}\bar{\nu}_{i}$ is induced by $Z^{0}$-penguins and
box diagrams, which, at the leading order, lead to the same effective operator. The
peculiarity of this process is that it is not affected by the QCD renormalization. This is
simply understood by noticing that the current ($\bar{s}\gamma^{\mu}P_{L}b$) is conserved in 
the limit of vanishing quark masses. Conserved currents have canonical dimensions and no
divergent counterterms arise. This ultraviolet behavior is not spoiled by consideration
of finite quark masses\cite{s15}.

The original electroweak sensitivity to the top mass is therefore preserved. On the other
hand, the experimental search for this rare B- transition is understandably much harder than
for the previous semileptonic decay.

For the e-neutrino and $\mu$-neutrino, we obtain the following result
\begin{eqnarray}
{\cal A}_{tot}(b\rightarrow s\nu_{i}\bar{\nu}_{i}(i=e, \mu))=A^{\nu}{\cal O}_{LLL}^{\nu} ,
\end{eqnarray}
where $ A^{\nu}=A^{Z}+B_{a}^{box}+B_{b}^{box} $ with $A^{Z}=\frac{e}{\sin\theta_{W} \cos\theta_{W}}\Gamma_{Z}$ and 
$B_{a,b}^{box}$ is given in Eq.\ (\ref{boxna}) and Eq.\ (\ref{boxnb}). The decay rate for the e, $\mu$-neutrino is 
then given by
\begin{eqnarray}
\Gamma(b\rightarrow s\nu_{i}\bar{\nu}_{i}(i=e, \mu))=\frac{m_{b}^{5}}{1536\pi^{3}}|A_{LLL}^{\nu}|^{2}.
\end{eqnarray}

As for the $\tau$-neutrino, the case is complicated because the $\tau$-neutrino mixed with
the neutralinos under our assumption. The interaction between the $Z_{0}$ boson and $\tau$-neutrino
can be written as:
\begin{eqnarray}
{\cal L}_{int}^{Z_{0}-\bar{\nu}_{\tau}-\nu_{\tau}} &=& -\frac{e}{2 \sin\theta_{W} \cos\theta_{W}}Z_{\mu}
\bar{\nu_{\tau}}\gamma^{\mu}P_{L}\nu_{\tau} \{ |Z_{N}^{3,1}|^{2} - |Z_{N}^{4,1}|^{2} + |Z_{N}^{5,1}|^{2} 
\} ,
\label{interznn}
\end{eqnarray}  
where $Z_{N}$ is the mixing matrix of $\tau$-neutrino and neutralinos, its definition can be found in the 
appendix. C. For the $\tau$-neutrino, we have
\begin{eqnarray}
{\cal A}_{tot}(b\rightarrow s\nu_{\tau}\bar{\nu}_{\tau})=F_{mix}A^{\nu}{\cal O}_{LLL}^{\nu}
\end{eqnarray}
with 
\begin{equation}
F_{mix} = |Z_{N}^{3,1}|^{2} - |Z_{N}^{4,1}|^{2} + |Z_{N}^{5,1}|^{2}.  \nonumber
\end{equation} 
The decay rate for the $\tau$-neutrino is given as:
\begin{eqnarray}
\Gamma(b\rightarrow s\nu_{\tau}\bar{\nu}_{\tau})=\frac{m_{b}^{5}}{1536\pi^{3}}F_{mix}^{2}|A_{LLL}^{\nu}|^{2},
\end{eqnarray}
here we neglect the mass of $\tau$-neutrino. We may now sum over all three types neutrino species:
\begin{equation}
\Gamma(b\rightarrow s\sum \nu_{i}\bar{\nu}_{i})=\frac{m_{b}^{5}}{1536\pi^{3}}(2 + 
F_{mix}^{2})|A_{LLL}^{\nu}|^{2} . \label{brsneu}
\end{equation}
As we shall see in the next section this leads to
a rate for $b \rightarrow s \sum \nu_{i} \bar{\nu_{i}}$ which is about four times larger than
the rate for $b \rightarrow s e^{+}e^{-}$.

\section{Numerical Results}

In this section, we will compute the branch ratio of the processes that have been analyzed in sect. III.
In the numerical evaluation below, we take $\alpha=\frac{e^{2}}{4\pi} \approx \frac{1}{128.8}$, $\alpha_{s}(M_{W})
\approx 0.118 $, $m_{e} \approx 0.511 {\rm MeV}$, $m_{\tau} \approx 1.777 {\rm GeV}$, $M_{Z} \approx 91.187
{\rm GeV}$,$M_{W} \approx 80.330 {\rm GeV}$, $m_{t} \approx 174. {\rm GeV}$, $m_{c} \approx 1.3 {\rm GeV}$ and $m_{b}
\approx 4.3 {\rm GeV}$.

In order to find out the allowed region in the parameter space, one has to 
take a number of constraint into account. First, we note that $m_{1}$, $m_{2}$, $\epsilon_{3}$,
$\mu$, $\tan\beta$, $\tan\theta_{\upsilon}$ and $l_{3}$ are the parameters that enter
into the chargino and neutralino mass matrices. The strongest constraint on
them follows from the fact that the $\tau$ mass has been experimentally measured
\cite{s20}, therefore, for any combination of those parameters, the lowest eigenvalue of 
Eq.\ (\ref{chaeig}) should agree with $m_{\tau}$.
 Also, $\nu_{\tau}$ has a laboratory upper limit of $24{\rm MeV}$ on its mass. The
two restrictions, together with the positive-definite condition
of the mass squared matrices, constrain the parameter space in a severe manner.

\begin{figure}
\setlength{\unitlength}{1mm}
\begin{picture}(180,190)(0,0)
\put(-30,-30){\includegraphics{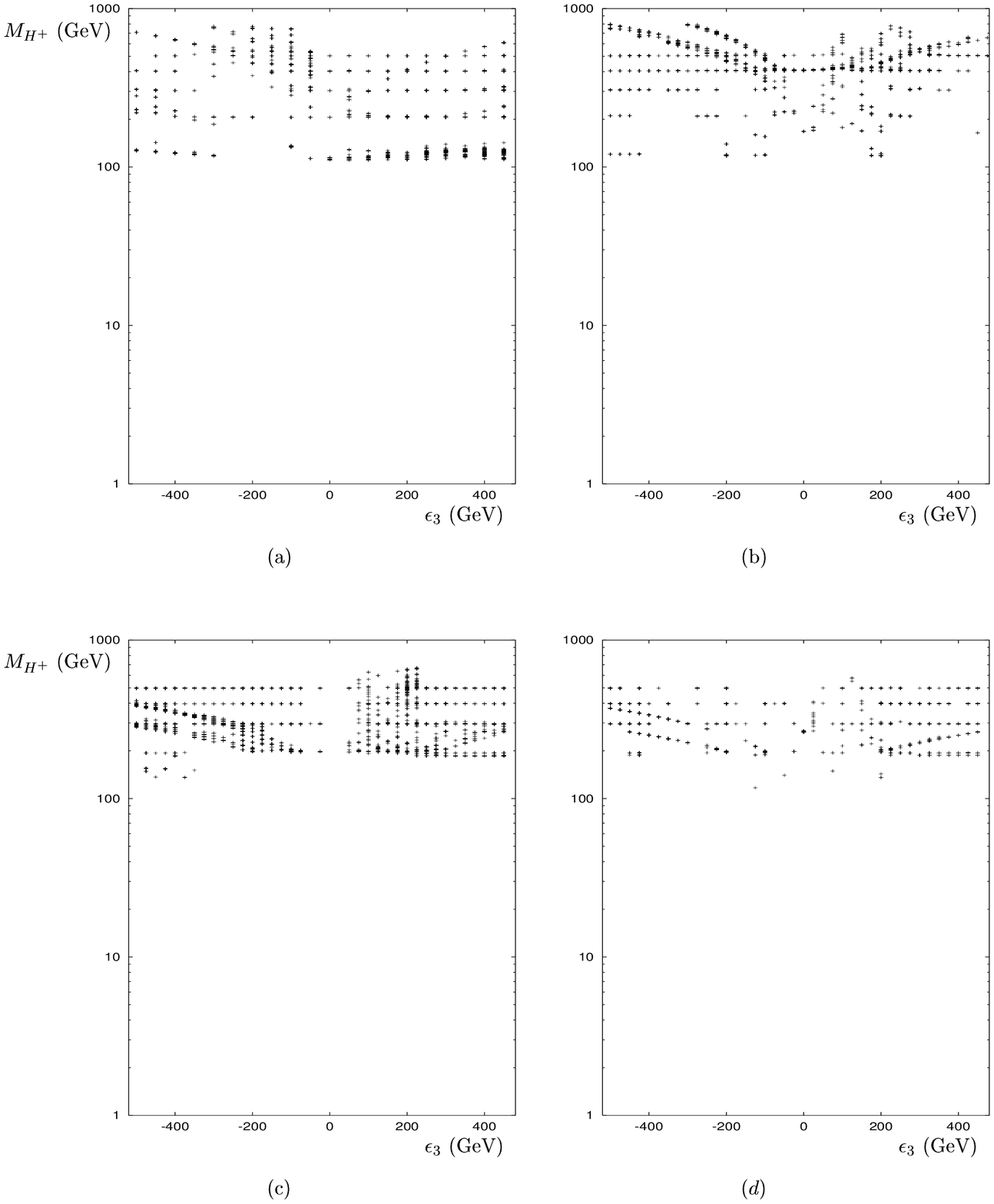}}
\end{picture}
\caption{Considering the experimental results of $b\rightarrow s+\gamma$, 
$b\rightarrow se^{+}e^{-}$, $M_{H^{+}}$ vary 
with $\epsilon_{3}$ when 
(a)$\tan\theta_{\upsilon}=20$, $\tan\beta=2$; (b)$\tan\theta_{\upsilon}=20$, $\tan\beta=40$;
(c)$\tan\theta_{\upsilon}=0.5$, $\tan\beta=2$; (d)$\tan\theta_{\upsilon}=0.5$, $\tan\beta=40$.
The other parameters are given in Eq.\ (\ref{range1}).}
\label{fig4}
\end{figure}

\begin{figure}
\setlength{\unitlength}{1mm}
\begin{picture}(180,190)(0,0)
\put(-30,-5){\includegraphics{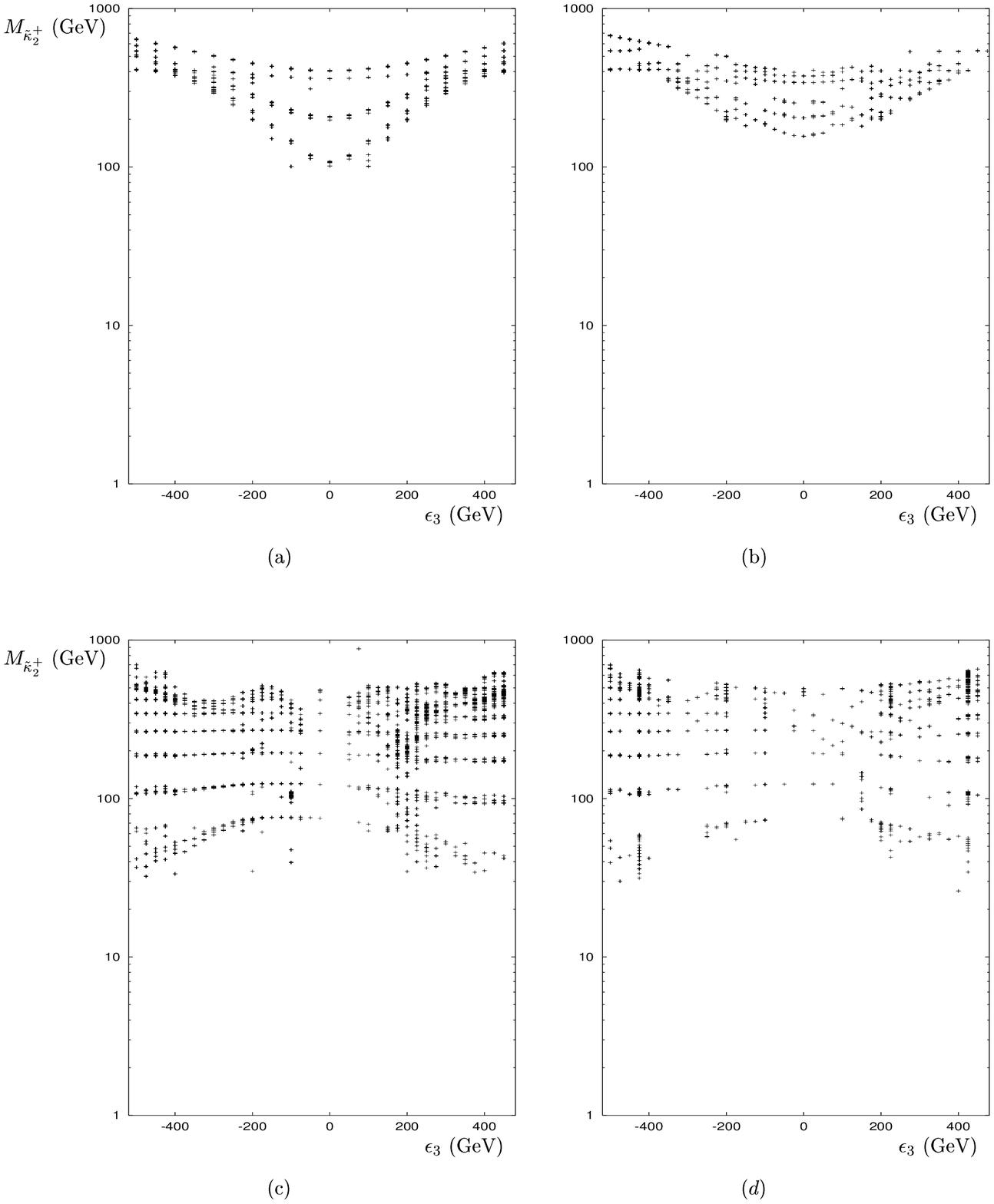}}
\end{picture}
\caption{Considering the experimental results of $b\rightarrow s+\gamma$, 
$b\rightarrow se^{+}e^{-}$, $m_{\kappa_{2}^{+}}$ vary 
with $\epsilon_{3}$ when 
(a)$\tan\theta_{\upsilon}=20$, $\tan\beta=2$; (b)$\tan\theta_{\upsilon}=20$, $\tan\beta=40$;
(c)$\tan\theta_{\upsilon}=0.5$, $\tan\beta=2$; (d)$\tan\theta_{\upsilon}=0.5$, $\tan\beta=40$.
The other parameters are given in Eq.\ (\ref{range1}).}
\label{fig5}
\end{figure}

We open our discussion by considering the experimental results of rare B processes
impact on the mass spectrum in the model. At present, the experimental bound on 
the $b \rightarrow s\gamma$, $b \rightarrow sl^{+}l^{-}$ are
$ 2 \times 10^{-4} \leq Br(b\rightarrow s+\gamma ) \leq 4.5 \times 10^{-4}$\cite{s20}
and $Br(b \rightarrow s e^{+}e^{-})^{CLEO} \leq 5.7 \times 10^{-5}$\cite{sadd1}
respectively. Not having the generality lost, we assume $l_{s3}=l_{3}$ in the
numerical calculation and the value of $l_{3}$ can be determined from $Det|m_{\tau}^{2}
- {\cal M}_{f}^{\dag}{\cal M}_{f}|=0$ when the relevant parameters are given. 
Furthermore, we interest the constraint on the mass spectrum that is imposed by
the experimental results of rare B processes, so we take the range of the parameters as: 
\begin{eqnarray}
&& -500 {\rm GeV} \leq B, B_{3} \leq 500 {\rm GeV}, \nonumber \\
&& 10^{4} {\rm GeV}^{2} \leq m_{R^{I}}^{2}, m_{Q^{I}}^{2}, m_{U^{I}}^{2}, m_{D^{I}}^{2} \leq 4 \times 
10^{6} {\rm GeV}^{2},  \nonumber \\
&& 100 {\rm GeV} \leq m_{1}, m_{2} \leq 1000 {\rm GeV}, \nonumber \\
&&-1000 {\rm GeV} \leq \mu \leq 1000 {\rm GeV}.
\label{range1}
\end{eqnarray} 
In the numerical program, the other parameters such as $\tan\beta$, $\tan\theta_{\upsilon}$
etc. are given in the figure caption. In Fig.\ \ref{fig4}, we plot the lightest
charged Higgs mass versus $\epsilon_{3}$ (in GeV). Under the bound of rare B processes,
we find the value of $M_{H^{+}}$ can vary from 100GeV to about 800GeV when the parameters
vary. Fig.\ \ref{fig5} shows the lightest chargino mass varies with the parameter
$\epsilon_{3}$, with other parameters taken as above. The point we should note is 
that $m_{\tilde{\kappa}_{2}^{+}}$ should be heavy when the $\tan\theta_{\upsilon}$
taken large value (such as $\tan\theta_{\upsilon} \sim 20$). When the $\tan\theta_{
\upsilon}$ taken small value, the mass of $\tilde{\kappa}_{2}^{+}$ can vary from
30GeV to several hundreds GeV.

\begin{figure}
\setlength{\unitlength}{1mm}
\begin{picture}(180,190)(0,0)
\put(-30,-20){\includegraphics{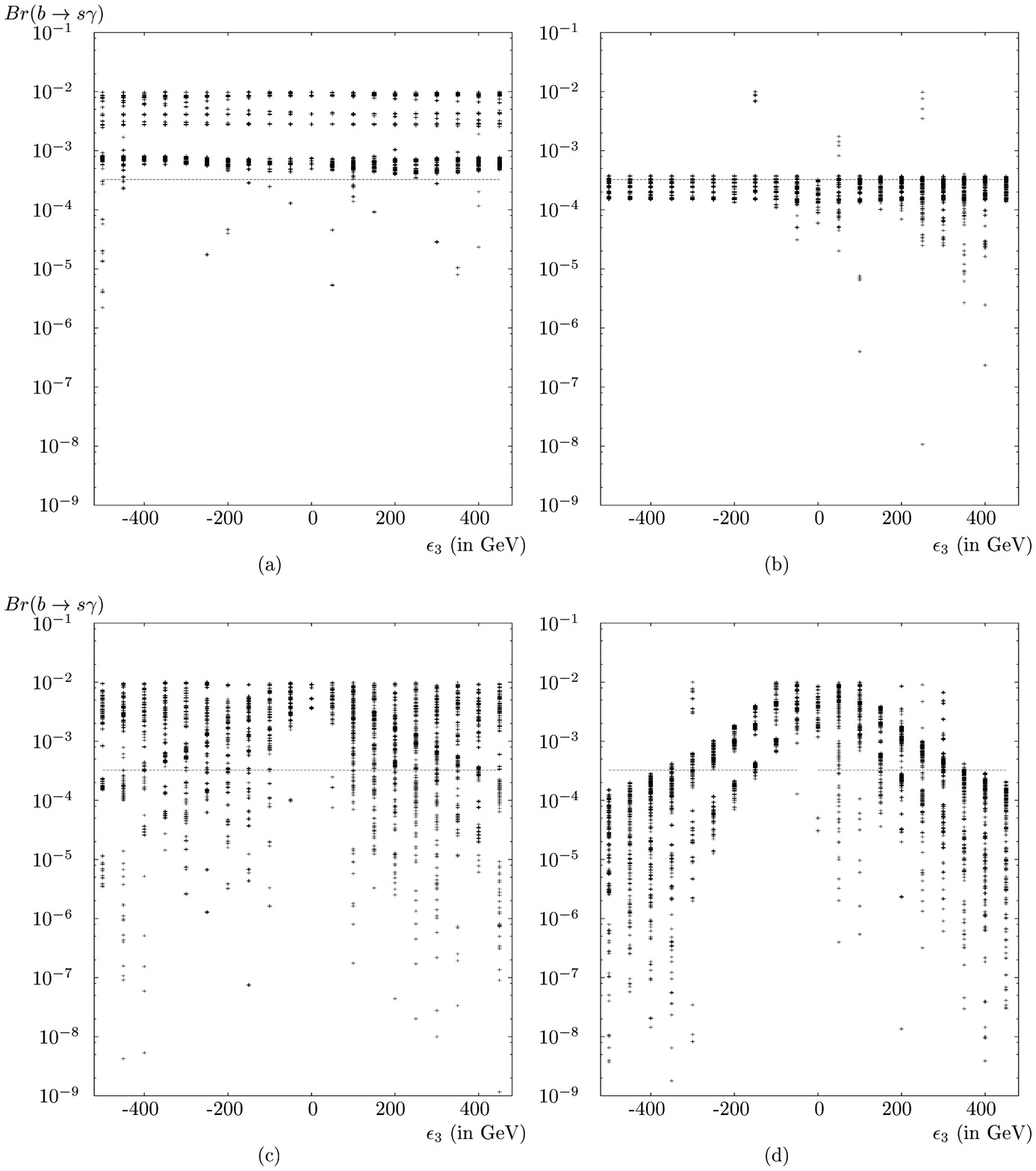}}
\end{picture}
\caption{Impose the Eq.\ (\ref{range2}), the branching ration of $b\rightarrow s+\gamma$ 
vary with $\epsilon_{3}$ at the $m_{b}$-scale when
(a)$\tan\theta_{\upsilon}=20$, $\tan\beta=2$;
(b)$\tan\theta_{\upsilon}=20$, $\tan\beta=40$;
(c)$\tan\theta_{\upsilon}=0.5$, $\tan\beta=2$;
(d)$\tan\theta_{\upsilon}=0.5$, $\tan\beta=40$.
The solid-lines are the predictions of SM.}
\label{fig6}
\end{figure}

Now, we turn to discuss the branch ratios of rare B processes. From Eq.\  (\ref{eq-21}),
we find the parameters $B$, $B_{3}$ enter the mass matrix of charged Higgs (just as in the 
mass matrices of CP-odd Higgs and CP-even Higgs) in forms $B\mu$ and $B_{3}\epsilon_{3}$.
Because we are interest in relatively light charginos and scalar particles, we take
\begin{eqnarray}
&& -100 {\rm GeV} \leq B \leq 100 {\rm GeV}, \nonumber \\
&& -100 {\rm GeV} \leq B_{3} \leq 100 {\rm GeV}, \nonumber \\
&& 10^{4} {\rm GeV}^{2} \leq m_{R^{I}}^{2}, m_{Q^{I}}^{2}, m_{U^{I}}^{2}, m_{D^{I}}^{2} \leq 2.5 \times 
10^{5} {\rm GeV}^{2},  \nonumber \\
&& 100 {\rm GeV} \leq m_{1}, m_{2} \leq 500 {\rm GeV}, \nonumber \\
&&-500 {\rm GeV} \leq \mu \leq 500 {\rm GeV}.
\label{range2}
\end{eqnarray} 
In Fig.\ \ref{fig6} we plot the branching ratio of $b\rightarrow s+\gamma $ 
as a function $\epsilon_{3} $ under some different value of $\tan\beta$ and $\tan\theta_
{\upsilon}$. The dependence on the remnant SUSY parameter such as $\mu$, $m_{1}$, $m_{2}$
is represented by the vertical width of the band. We see that positive interference with 
the different sources of SUSY contributions can produce an intensive enhancement over the QCD-
corrected SM prediction (horizontal solid line) when those parameters are assigned suitable
values. Sometimes, the supersymmetric contributions dominate over the SM contributions
resulting in significant deviation from the SM prediction. 

\begin{figure}
\setlength{\unitlength}{1mm}
\begin{picture}(180,190)(0,0)
\put(-30,-20){\includegraphics{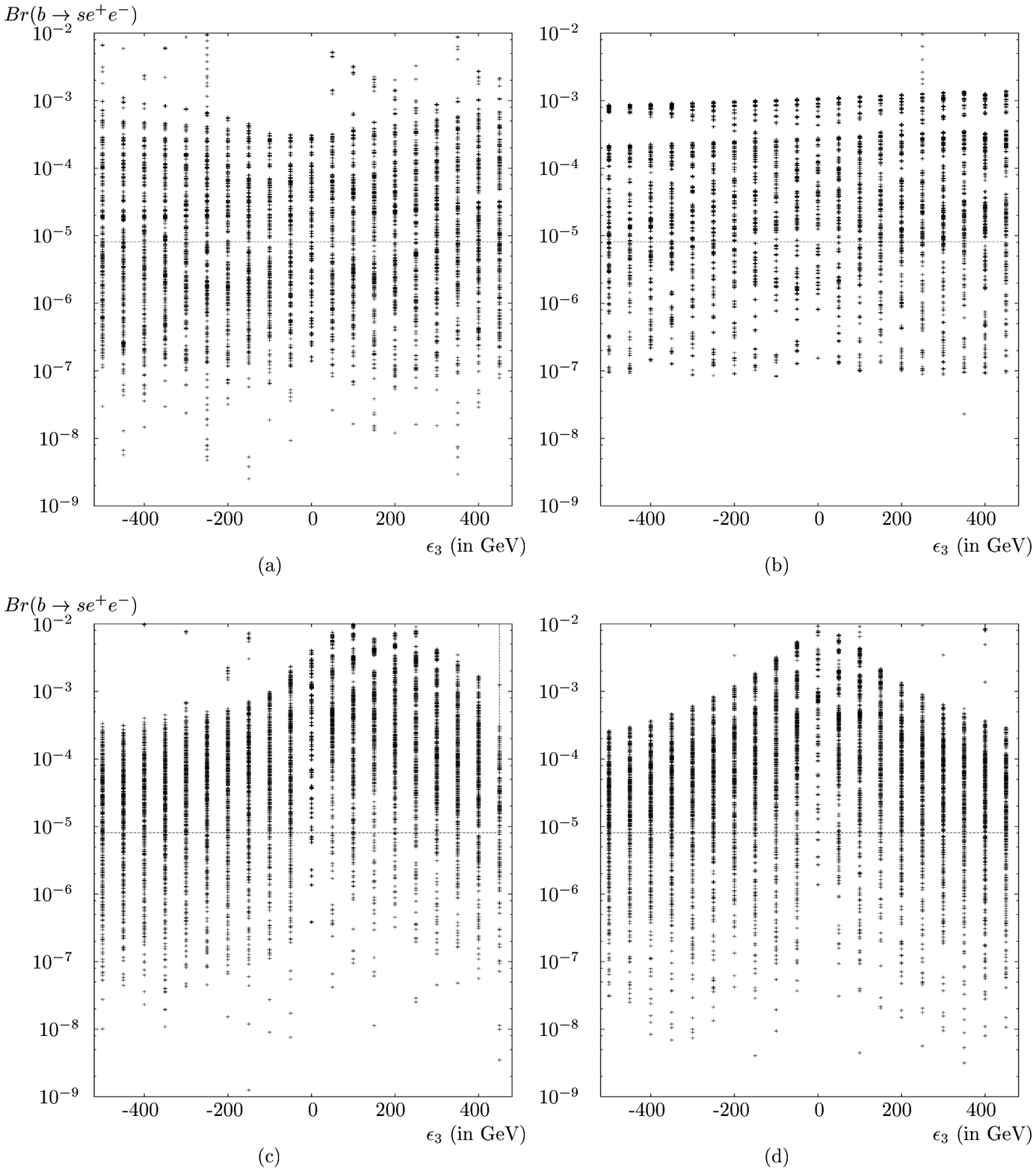}}
\end{picture}
\caption{Impose the Eq.\ (\ref{range2}), the branching ration of $b\rightarrow s+e^{+}e^{-}$ 
vary with $\epsilon_{3}$ at the $m_{b}$-scale when
(a)$\tan\theta_{\upsilon}=20$, $\tan\beta=2$;
(b)$\tan\theta_{\upsilon}=20$, $\tan\beta=40$;
(c)$\tan\theta_{\upsilon}=0.5$, $\tan\beta=2$;
(d)$\tan\theta_{\upsilon}=0.5$, $\tan\beta=40$.
The solid-lines are the predictions of SM.}
\label{fig7}
\end{figure}

\begin{figure}
\setlength{\unitlength}{1mm}
\begin{picture}(180,190)(0,0)
\put(-30,-20){\includegraphics{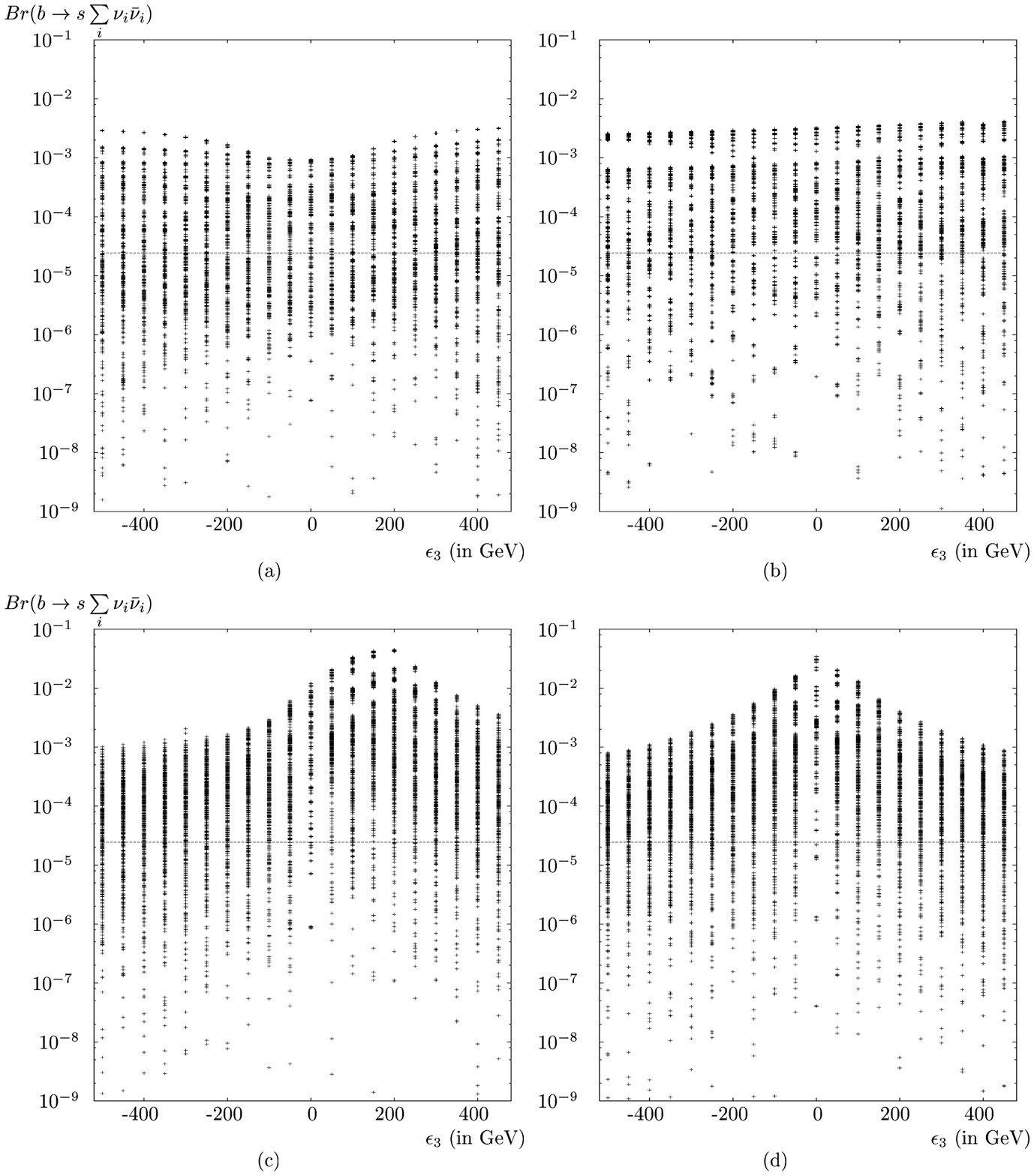}}
\end{picture}
\caption{Impose the Eq.\ (\ref{range2}), the branching ration of $b\rightarrow 
s+\sum\limits_{i}\nu_{i}\bar{\nu}_{i}$ vary with $\epsilon_{3}$ at the $m_{b}$-scale when
(a)$\tan\theta_{\upsilon}=20$, $\tan\beta=2$;
(b)$\tan\theta_{\upsilon}=20$, $\tan\beta=40$;
(c)$\tan\theta_{\upsilon}=0.5$, $\tan\beta=2$;
(d)$\tan\theta_{\upsilon}=0.5$, $\tan\beta=40$.
The solid-lines are the predictions of SM.}
\label{fig8}
\end{figure}

Let us discuss the semileptonic FCNC decay $b\rightarrow s+ e^{+}e^{-}$ and $b\rightarrow 
s \sum\nu_{i}\bar{\nu}_{i}$ numerically. In the SM, the QCD-corrected $Br(b\rightarrow s+e^{+}e^{-})$ is about
$9 \times 10^{-6}$. The addition of the one-loop contributions where SUSY particles are 
present modify the prediction up to about four times the SM prediction as it can be gathered
from Fig.\ \ref{fig7}. The CLEO collaboration has already been searched
for inclusive $b \rightarrow s l^{+}l^{-}(l=e, \mu)$. The results are\cite{sadd1}:
\begin{eqnarray}
Br(b \rightarrow s e^{+}e^{-})^{CLEO} \leq 5.7 \times 10^{-5}, \nonumber \\
Br(b \rightarrow s\mu^{+}\mu^{-})^{CLEO} \leq 5.8 \times 10^{-5}. \nonumber \\
\end{eqnarray}
By comparing the numerical result with experiment, we find that we can not excluded the
large value of $\epsilon_{3}$. Analogous
considerations hold for $b\rightarrow s\sum\nu_{i}\bar{\nu}_{i}$ (Fig.\ \ref{fig8}). 
The R-breaking terms can have an
appreciable effect for the $b \rightarrow s\sum \nu_{i}\bar{\nu}_{i}$. This is the main difference
between the BRPV model and the usual SUSY model with R-parity.

In summary, as a simple extension of the MSSM which introduce R-parity violation, the R-breaking
terms in BRPV model can give an appreciable effect for the rare B-processes. From the present
experimental bound on those processes, we can get some constraint on the mass spectrum in this model
under some suitable assumptions.

\vspace{2cm}
{\Large\bf Acknowledgment} This work was supported in part by the National Natural
Science Foundation of China and the Grant No. LWLZ-1298 of the Chinese Academy of
Sciences.

\appendix
\section{The defintion of various functions}

We collect in this appendix the various functions that were used in the text.
The one-variable functions obtained from the penguin diagrams are given as:
\begin{eqnarray}
&& f_{2}^{(0)}(x)=\frac{x}{1-x}\ln x , \nonumber \\
&& f_{3a}^{(0)}(x) = -\frac{1}{1-x}\{1+\frac{x}{1-x}\ln x \} , \nonumber \\
&& f_{3b}^{(0)}(x) = \frac{1}{1-x}\{1+\frac{1}{1-x}\ln x \} , \nonumber \\
&& f_{3a}^{(1)}(x)=\frac{2\ln x}{1-x}+\ln x - \frac{\ln x}{(1-x)^{2}} - \frac{1}{(1-x)} , \nonumber \\
&& f_{3b}^{(1)}(x)=\frac{2x\ln x}{1-x}-\ln x + \frac{x^{2}\ln x}{(1-x)^{2}} + \frac{x}{(1-x)} , \nonumber \\
&& f_{4a}^{(1)}(x) = \frac{1}{(x-1)}[\frac{1}{2} + \frac{x}{x-1} - \frac{x^{2}}{(x-1)^{2}}\ln x ] , \nonumber \\
&& f_{4b}^{(1)}(x) = \frac{2}{(x-1)}[\frac{1}{2} - \frac{x}{x-1} - \frac{x}{(x-1)^{2}}\ln x ] , \nonumber \\
&& f_{4c}^{(1)}(x) = -\frac{1}{(x-1)}[\frac{1}{2} - \frac{1}{x-1} - \frac{1}{(x-1)^{2}}\ln x ] , \nonumber \\
&& f_{5a}^{(2)}(x) = \frac{1}{3(x-1)}+\frac{x}{2(x-1)^{2}}+\frac{x^{2}}{(x-1)^{3}}-
\frac{x^{3}}{(x-1)^{4}}\ln x , \nonumber \\
&& f_{5b}^{(2)}(x) = \frac{5}{2(x-1)}-\frac{3x}{2(x-1)^{2}}+\frac{3x^{2}}{(x-1)^{3}}-
\frac{3x^{2}}{(x-1)^{4}}\ln x , \nonumber \\
&& f_{5c}^{(2)}(x) = -\frac{1}{2(x-1)}+\frac{3x}{2(x-1)^{2}}+\frac{3}{(x-1)^{3}}-
\frac{3x}{(x-1)^{4}}\ln x ,  \nonumber \\
&& f_{5d}^{(2)}(x) = -\frac{1}{3(x-1)}+\frac{1}{2(x-1)^{2}}-\frac{1}{(x-1)^{3}}+\frac{1}{(x-1)^{4}}\ln x .
\end{eqnarray}
The two- and three-variable functions obtained from penguin and box diagrams are
\begin{eqnarray}
&& f_{3c}^{(0)}(x,y) = -\frac{1}{x-y}[\frac{x}{x-1}\ln x - \frac{y}{y-1}\ln y ] ,  \nonumber \\
&& f_{3c}^{(1)}(x,y) = -\frac{1}{2(x-y)}[\frac{x^{2}}{x-1}\ln x - 
\frac{y^{2}}{y-1}\ln y ] ,  \nonumber \\
&& f_{4d}^{(1)}(x,y) = \frac{1}{x-y}[\frac{x^{2}}{(x-1)^{2}}\ln x - 
\frac{1}{x-1} - (x \rightarrow y)] ,  \nonumber \\
&& f_{4e}^{(1)}(x,y,z) = \frac{1}{x-y}\{ \frac{1}{x-z}[\frac{x^{2}}{(x-1)}\ln x - \frac{3x}{2} - 
(x \rightarrow z)] -(x \rightarrow y ) \} .
\end{eqnarray}

\section{The relevant Feynman Rules in SUSY model with bilinear R-parity violation}

In this appendix, we give some relevant Feynman rules in the supersymmetric model with bilinear
R-parity violation that were used in the paper.

\begin{figure}
\setlength{\unitlength}{1mm}
\begin{picture}(180,200)(0,0)
\put(-30,-40){\includegraphics{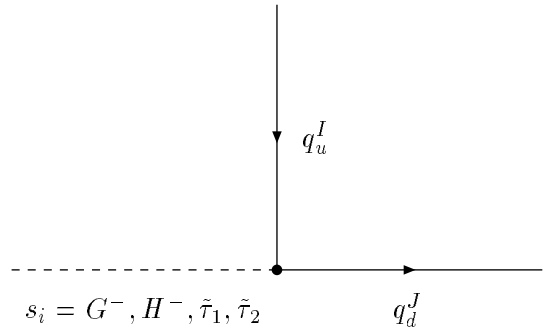}}
\end{picture}
\caption{The coupling between $s_{i}=(G^{-}$, $H^{-}$, $\tilde{\tau}^{-}_{1}$, $\tilde{\tau}^{-}_{1})$ and 
$q_{u}^{I}$, $q_{d}^{J}$}
\label{fig10}
\end{figure}

\begin{figure}
\setlength{\unitlength}{1mm}
\begin{picture}(180,200)(0,0)
\put(-30,-40){\includegraphics{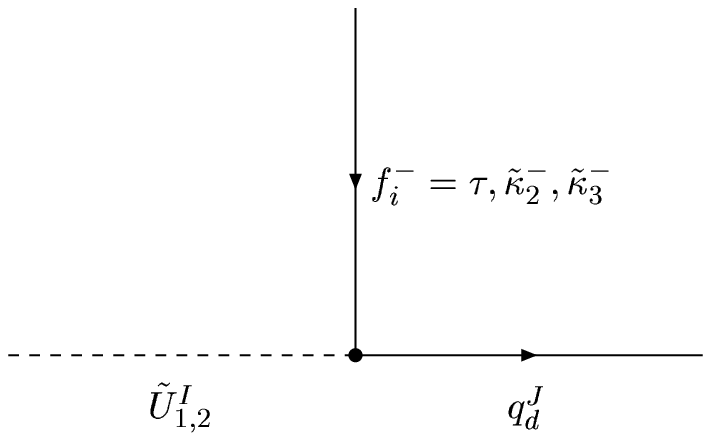}}
\end{picture}
\caption{The coupling between $f_{i}=(\tau^{-}$, $\tilde{\kappa}^{-}_{2}$, 
$\tilde{\kappa}^{-}_{3})$ and $q_{d}^{J}$, $\tilde{U}^{I}_{j}$ }
\label{fig11}
\end{figure}

\begin{figure}
\setlength{\unitlength}{1mm}
\begin{picture}(180,200)(0,0)
\put(-30,-40){\includegraphics{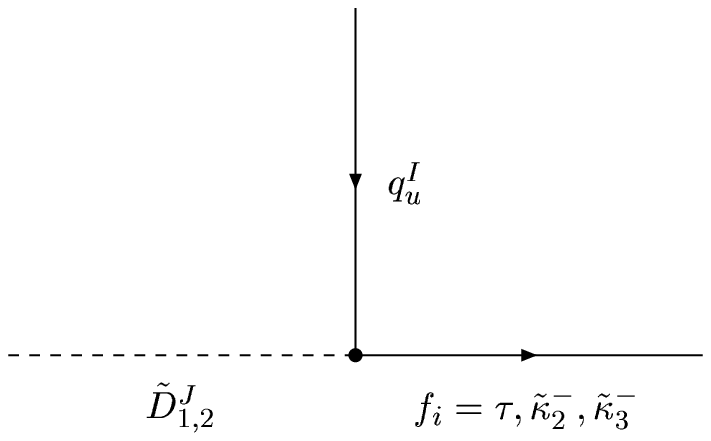}}
\end{picture}
\caption{The coupling between $f_{i}=(\tau^{-},\tilde{\kappa}^{-}_{2},
\tilde{\kappa}^{-}_{3})$ and $q_{u}^{I}$,$\tilde{D}^{J}_{j}$ }
\label{fig12}
\end{figure}

\begin{figure}
\setlength{\unitlength}{1mm}
\begin{picture}(180,200)(0,0)
\put(-30,-40){\includegraphics{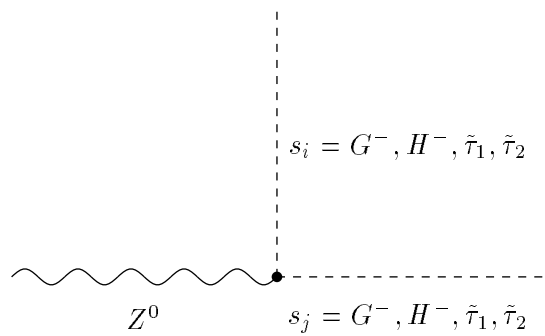}}
\end{picture}
\caption{The coupling between $Z^{0}$ and $s_{i}=(G^{-},H^{-},\tilde{\tau}_{1},\tilde{\tau}_{1})$}
\label{fig13}
\end{figure}

It is first convenient to introduce the mixing matrices relative to the scalar-quark sector. We
denote with $\tilde{Q}^{I}_{L,R}$ the squark current eigenstates (where $I=1$, $2$, $3$ is the generation
label and $Q=U$, $D$), and with $\tilde{Q}^{I}_{1,2}$ the corresponding mass eigenstates of the I-th
generation (we neglect the generation-mixing of the scalar-quark and scalar-lepton). The $2\times 2$
mixing matrices $Z_{\tilde{Q}^{I}}$ are defined by
\begin{eqnarray}
\tilde{Q}^{I}_{L} &=& Z_{\tilde{Q}^{I}}^{1,i}\tilde{Q}^{I}_{i}  , \nonumber  \\
\tilde{Q}^{I}_{R} &=& Z_{\tilde{Q}^{I}}^{2,i}\tilde{Q}^{I}_{i} .  
\end{eqnarray} 
Similar, the $2\times 2$ mixing matrices $Z_{\tilde{E}^{I}}$ are defined as
\begin{eqnarray}
\tilde{L}^{I} &=& Z_{\tilde{E}^{I}}^{1,i}\tilde{E}^{I}_{i} , \nonumber  \\
\tilde{R}^{I*} &=& Z_{\tilde{E}^{I}}^{2,i}\tilde{E}^{I}_{i} 
\end{eqnarray} 
with $\tilde{L}^{I}$, $\tilde{R}^{I}$ are the slepton current eigenstates and $\tilde{E}_{1,2}^{I}$ are 
the corresponding mass eigenstates. 
The $S_{i}$, ($S_{i=1,2,3,4}=G^{-}$, $H^{-}$, $\tilde{\tau}_{1}$, $\tilde{\tau}_{2}$)
up quark and down quark couplings can be written as (Fig.\ \ref{fig10})
\begin{equation}
i[\frac{em_{d}^{I}}{\sqrt{2}M_{W}\sin\theta_{W}\sin\theta_{v}\cos\beta}Z_{H}^{1i}P_{L} +
\frac{em_{u}^{J}}{\sqrt{2}M_{W}\sin\theta_{W}\sin\theta_{v}\sin\beta}Z_{H}^{2i}P_{R}]C_{IJ} .
\end{equation}
The couplings of down quark, up scalar quark and chargino are given by (Fig.\ \ref{fig11})
\begin{eqnarray}
&& i[(\frac{-e}{\sin\theta_{W}}Z_{\tilde{U}^{I}}^{1i*}Z_{1j}^{+}+\frac{em_{u}^{J}}{\sqrt{2}M_{W}\sin\theta_{W}
\sin\theta_{v}\sin\beta}Z_{\tilde{U}^{I}}^{2i*}Z_{2j}^{+})P_{L}+ \nonumber \\
&& \frac{em_{d}^{I}}{\sqrt{2}M_{W}\sin\theta_{W}
\sin\theta_{v}\cos\beta}Z_{\tilde{U}^{J}}^{1i*}Z_{2j}^{-}P_{R}]C_{IJ}^{*} .
\end{eqnarray}
The couplings of up quark, down scalar quark and chargino are given by (Fig.\ \ref{fig12})
\begin{eqnarray}
&& i[(\frac{-e}{\sin\theta_{W}}Z_{\tilde{D}^{I}}^{1i}Z_{1j}^{+}+\frac{em_{d}^{I}}{\sqrt{2}M_{W}\sin\theta_{W}
\sin\theta_{v}\cos\beta}Z_{\tilde{D}^{I}}^{2i}Z_{2j}^{-})P_{L}+ \nonumber \\
&& \frac{em_{u}^{J}}{\sqrt{2}M_{W}\sin\theta_{W}
\sin\theta_{v}\sin\beta}Z_{\tilde{D}^{I}}^{1i}Z_{2j}^{+*}P_{R}]C_{IJ}^{*} .
\end{eqnarray}
The couplings of $Z^{0}$ and $S_{i}$ are (Fig.\ \ref{fig13})
\begin{eqnarray}
i\frac{e}{2\sin\theta_{W}\cos\theta_{W}}[(\cos^{2}\theta_{W}-\sin^{2}\theta_{W})\delta{ij}-Z_{H}^{4i}
Z_{H}^{4j*}](p+k)^{\mu} ,
\end{eqnarray}
where $S_{i}=(G^{-}$, $H^{-}$, $\tilde{\tau}_{1}$, $\tilde{\tau}_{2})$, $i=1$, $2$, $3$, $4$. The
matrices $Z_{H}$, $Z^{\pm}$ have been defined as above.

\section{The mixing between $\tau$-neutrino and neutralinos}

In this appendix, we give the mass matrix of $\tau$-neutrino and neutralinos under our
assumption. 

From Eq.\ (\ref{supernew}) and Eq.\ (\ref{softnew}), the $\tau$ neutrino-neutralino
mass terms in the Lagrangian are
\begin{equation}
{\cal L}_{neutralino} = -\frac{1}{2}(\Psi^{0})^{T}{\cal M}_{N}\Psi^{0} + h.c.
\end{equation}
with $(\Psi^{0})^{T} = (-i\lambda_{B}$, $-i\lambda_{A}^{3}$, $\psi_{H^{1}}^{1}$, $\psi_{H^{2}}^{2}$,
$\nu_{\tau_{L}} )$
and 
\begin{equation}
{\cal M}_{N} =
\left( \begin{array}{ccccc}
2m_{1} & 0 & -\frac{1}{2}g'\upsilon_{1} & \frac{1}{2}g'\upsilon_{2} & -\frac{1}{2}g'\upsilon_{3}  \\
0 & 2m_{2} & \frac{1}{2}g\upsilon_{1} & -\frac{1}{2}g\upsilon_{2} & \frac{1}{2}g\upsilon_{3}  \\
-\frac{1}{2}g'\upsilon_{1} &  \frac{1}{2}g\upsilon_{1} & 0 & -\frac{1}{2}\mu & 0 \\
\frac{1}{2}g'\upsilon_{2}  &  -\frac{1}{2}g\upsilon_{2} & -\frac{1}{2}\mu & 0 & \frac{1}{2} \epsilon_{3} \\
-\frac{1}{2}g'\upsilon_{3} & \frac{1}{2}g\upsilon_{3} & 0 & \frac{1}{2}\epsilon_{3} & 0
\end{array}  \right)
\end{equation}
The formulae of mixing matrix are:
\begin{eqnarray}
-i\lambda_{B} &=& Z_{N}^{1,i} \chi_{i}^{0},  \nonumber \\
-i\lambda_{A}^{3} &=& Z_{N}^{2,i} \chi_{i}^{0}, \nonumber \\
\psi_{H^{1}}^{1} &=& Z_{N}^{3,i} \chi_{i}^{0},  \nonumber  \\
\psi_{H^{2}}^{2} &=& Z_{N}^{4,i} \chi_{i}^{0},  \nonumber  \\
\nu_{\tau_{L}} &=& Z_{N}^{5,i} \chi_{i}^{0}
\end{eqnarray}
and
\begin{equation}
\nu_{\tau} = 
\left(  
\begin{array}{c}
\chi_{1}^{0}  \\
\bar{\chi}_{1}^{0}
\end{array}  \right) 
\end{equation} 

\begin{equation}
\kappa_{i}^{0} = 
\left(  
\begin{array}{c}
\chi_{i+1}^{0}  \\
\bar{\chi}_{i+1}^{0}
\end{array}  \right)  
\end{equation}
with $i=1$, 2, 3, 4. Here, we identify the $\nu_{\tau}$ as the lightest mass eigenstate of
the mass matrix. The matrix $Z_{N}$ satisfies the following condition: 
$Z_{N}^{T} {\cal M}_{N} Z_{N} = diag(m_{\nu_{\tau}}$, $m_{\kappa_{1}^{0}}$, $m_{\kappa_{2}^{0}}$, 
$m_{\kappa_{3}^{0}}$, $m_{\kappa_{4}^{0}})$. 
Similar to the mixing matrices of $\tau$-chargino sector, we can assume $m_{\nu_{\tau}}$, $m_{
\kappa_{i}^{0}}$ $(i=1$, $2$, $3$, $4)$
positive and $m_{\kappa_{4}^{0}} > m_{\kappa_{3}^{0}} > m_{\kappa_{2}^{0}} > m_{\kappa_{1}^{0}} 
> m_{\nu_{\tau}}$.

\end{document}